\documentclass{article}

\usepackage{graphics}
\usepackage{graphicx}
\usepackage{amsthm} 
\usepackage{amsbsy} 
\usepackage{amsmath,amsfonts,amssymb,times}
\usepackage{dsfont}
\usepackage{hyperref}
\usepackage{natbib}
\usepackage{pstricks}
\usepackage{authblk}
\usepackage{comment}

\usepackage{amsmath} 
\usepackage{amsthm} 
\usepackage{amssymb} 
\usepackage{amsbsy} 
\usepackage{graphicx} 
\usepackage{color} 
\usepackage{enumerate} 
\usepackage{todonotes} 
\usepackage{tocloft} 
\usepackage{comment} 

\newcommand{\bB}{\begin{color}{blue}}
\newcommand{\bR}{\begin{color}{red}}
\newcommand{\bM}{\begin{color}{magenta}}
\newcommand{\bC}{\begin{color}{cyan}}
\newcommand{\bW}{\begin{color}{white}}
\newcommand{\bG}{\begin{color}{green}}
\newcommand{\bY}{\begin{color}{yellow}}
\newcommand{\e}{\end{color}}

\definecolor{red}{rgb}{0.9,0,0}
\definecolor{green}{rgb}{0,0.8,0}
\definecolor{blue}{rgb}{0,0,0.8}
\definecolor{cautionred}{rgb}{1.0,0,0}
\definecolor{maroon}{rgb}{0.7,0,0}
\definecolor{ngreen}{rgb}{0.3,0.7,0.3}
\definecolor{golden}{rgb}{0.8,0.6,0.1}

\newcommand{\citeapos}[1]{\citeauthor{#1}'s (\citeyear{#1})}

\newlistof{ideaTagCounter}{ideaTags}{List of paragraph notes} 
\newcommand{\parnote}[1]{
	\refstepcounter{ideaTagCounter} 
	\addcontentsline{ideaTags}{ideaTagCounter}{
		(\theideaTagCounter)\quad\thesubsection \quad#1
		}
	\todo[linecolor=blue!40!white, backgroundcolor=blue!20!white,
		bordercolor=white, size=\footnotesize]{#1}
}
\reversemarginpar

\theoremstyle{remark}
\theoremstyle{definition}
\newcommand{\beq}{\begin{equation}}
\newcommand{\eeq}{\end{equation}}
\newcommand{\ba}{\begin{eqnarray}}
\newcommand{\ea}{\end{eqnarray}}
\newcommand{\ban}{\begin{eqnarray*}}
\newcommand{\ean}{\end{eqnarray*}}
\newcommand{\bit}{\begin{itemize}}
\newcommand{\eit}{\end{itemize}}
\newcommand{\ben}{\begin{enumerate}}
\newcommand{\een}{\end{enumerate}}
\newcommand{\bqu}{\begin{quote}}
\newcommand{\equ}{\end{quote}}

\newcommand{\Hilb}{\mathbf{Hilb}}
\newcommand{\nCob}{\boldsymbol{n}\mathbf{Cob}}

\newcommand{\Set}{\mathbf{Set}}
\newcommand{\Rel}{\mathbf{Rel}}

\title{Categorical Generalization and Physical Structuralism}

\begin{document}

\author[1]{Raymond Lal\thanks{raymond.lal@cs.ox.ac.uk}}
\author[2]{Nicholas J. Teh\thanks{njywt2@cam.ac.uk}}
\affil[1]{Quantum Group, Department of Computer Science \newline University of Oxford}
\affil[2]{DAMTP and Department of History and Philosophy of Science \newline University of Cambridge}

\maketitle
\begin{abstract}
Category theory has become central to certain aspects of theoretical physics. 
\cite{Bain2011}  has recently argued that this has significance for ontic structural realism.
We argue against this claim.
In so doing, we uncover two pervasive forms of category-theoretic generalization.
We call these `generalization by duality' and `generalization by categorifying  physical processes'.
We describe in detail how these arise, and explain their significance using detailed examples.
We show that their significance  is two-fold: the articulation of high-level physical concepts, and the generation of new models.
\end{abstract} 
\tableofcontents

\section{Introduction}

Recently, \cite{Bain2011} has offered an argument in defense of radical ontic structural realism (ROSR) which draws on category-theoretic resources. His argument is largely based on the idea that category theory offers the possibility of formulating physical theories at an especially high level of generality; this level of generality is in turn connected to the elimination of objects from physical theories.

In this paper, we will first argue that Bain's argument fails.\footnote{N.B. Contemporaneously with our paper, \cite{Lam2013} have offered a discussion of Bain's argument. We agree with many of their criticisms, but for the most part, they leave the deeply category-theoretic aspects of his argument---especially the connection with generalisation---undiscussed. This is our main topic, and as such, there is little overlap between our work and theirs.} Our analysis of Bain's argument will show that it employs two distinct strategies, each of which implicitly rests on a form of generalization that involves category theory. Indeed, it turns out that these are the two main forms of category-theoretic generalization that have been hitherto invoked in applications of category theory to physics.

This in turn paves the way for the more positive task of our paper. We shall extract the two forms of generalization from Bain's argument and develop each in turn, thus providing a conceptual articulation of the two forms of generalization, and an analysis of the role that each plays in the physical examples invoked by Bain. 

This task is easy to motivate, as the general question of how one can generalize physical theories is of immense interest to the philosophy of physics; indeed, it would be a great discovery if it turned out that some very comprehensive and systematic means existed. 
In our investigation, we shall be concerned specifically with textit{category-theoretic} types of generalization within physics.
An advantage of our approach is that, rather than approaching the question of generalizing physical theories with a preconceived and overly abstract idea of what `generalization' amounts to, our approach has the benefit of beginning with present applications of category theory to physics (indeed, the ones suggested by Bain's argument) and extracting from these two forms of generalization that are grounded in the actual practice of physics. 
Furthermore, our results provide a partial explanation of how category theory can be useful for physics.

We now proceed to give a prospectus of the paper. Section 2 covers background material that will be assumed in the rest of the paper: Section 2.1 reviews several key notions from category theory, and Section 2.2 highlights the way in which category theory is uncontroversially taken to be `structuralist' by its practitioners, independently of its potential connection with any philosophical doctrines concerning `structural realism'.

Section 3 points out that there are really two different argumentative strategies at work in \cite{Bain2011} and analyzes each in turn. Section 3.1 discusses the first strategy, which involves what we call `generalization by duality' (GenDual), and Section 3.2 discusses the second strategy, which involves what we call `generalization by categorification' (GenCat).

By our lights, both of these strategies are unsuccessful. However, their structure is of independent interest, because they rest on important intuitions about how category theory helps us to generalize physical theories.
Section 4 develops these intuitions and clarifies exactly how these two forms of generalization are supposed to work.
Section 4.1.1 develops the framework for generalization by duality (GenDual) and Section 4.1.2 proceeds to consider the particular form of (GenDual) that is invoked in Bain's Einstein Algebra example. We shall argue that the details of how (GenDual) is applied do not work in Bain's favor. Similarly, Section 4.2.1 develops the framework for generalization by categorification (GenCat) and Section 4.2.2 discusses its putative application in TQFTs.
Finally, Section 4.3 discusses the differences between (GenDual) and (GenCat), and the role that category theory plays in each of these forms of generalization.

\section{Categories and structuralism}\label{sec:CTandSR}

This section connects category theory with various themes that we will touch upon in the paper.
Section 2.1 introduces the basic notions of category theory that will be required in the rest of the paper. 
Section 2.2 then proceeds to sketch the simple `structuralist' idea that is implicit (and indeed uncontroversial) in the mathematical use of category theory. We contrast this with the metaphysical doctrine of ontic structural realism, since subsequent sections will discuss whether category theory can furnish a defense of ontic structural realism. 
\subsection{Categories: abstract and concrete}\label{sec:concrete}

For the reader's convenience, we begin by providing the definition of a category. 

A \em category \em consists of: a class of \em objects\em, denoted $A,B,\dots$, and a class of morphisms, denoted $f,g,\dots$, which satisfy the following conditions.
A morphism $f$ is written as $f:A\rightarrow B$, meaning that it has the object $A$ as its \em domain\em, and the object $B$ as its codomain.
There is a \em composition law \em $-\circ-$ for morphisms, which means that for any two morphisms $f:A\rightarrow B$ and $g:C\rightarrow D$, there is a morphism $g\circ f$ whenever $B=C$ (which is referred to as `type-matching').
This composition law is associative, so that $h\circ (g\circ f)=(h\circ g)\circ f$ for all morphisms with appropriate type-matching.
For every object $A$, there is always an \em identity morphism \em $1_A$, so that for any morphism $f:A\rightarrow B$, the equation $1_B\circ f = f= f\circ 1_A$ is satisfied. 

Notice that in this definition, one does not need to further specify the details of the objects and morphisms of a category. For all we know, the objects might not even be sets, and the morphisms might not even be functions. If, however, the objects of the category are sets that are equipped with some additional structure, then category theorists say that the category is \textit{concrete}.\footnote{More formally, a concrete category is defined as a pair $(C, F)$ where $C$ is a category and $F$ is a faithful functor from $C$ to \textbf{Set}. 
Note that there is a different use of `concrete category' in some of the philosophy of mathematics literature \citep{Landry2013}, where it means `a category that provides an interpretation of the Eilenberg-MacLane axioms'. 
This is not the sense that we have in mind. 
Instead our usage follows that of the major category theory textbooks, see e.g.~\cite[p.~26]{MacLane2000} or \cite[p.~4]{Lambek1988}.}

Here are some examples of familiar concrete categories (some of which will figure in the discussion below):

\begin{itemize}
\item The category $\mathbf{Set}$, whose objects are sets and whose morphisms are functions between sets.
 
\item The category $\mathbf{Grp}$, whose objects are groups and whose morphisms are homomorphisms between groups.

\item The category $\mathbf{Top}$, whose objects are topological spaces and whose morphisms are continuous maps between the spaces.

\item The category $\mathbf{Man}$, whose objects are smooth manifolds and whose morphisms are smooth maps between the manifolds.
\end{itemize}

Any category comes equipped with a way of saying that two objects of the category are `structurally the same', viz. the concept of an \textit{iso}-morphism. This is in fact a generalization of the usual set-theoretic definition of isomorphism, i.e. a bijective structure-preserving map between two structured sets. By contrast, category theory offers a definition purely in terms of \textit{morphisms}, and does not explicitly quantify over the elements of the objects:
\begin{quote}
Two objects $A$, $B$ in a category $\mathcal{C}$ are \textit{isomorphic} iff there exist morphisms $f: A \rightarrow B$ and $g: B \rightarrow A$ such that $ g\circ f = 1_A$ and $f\circ g = 1_B$.
\end{quote}
For example, an isomorphism in $\mathbf{Top}$ is a homeomorphism, and an isomorphism in $\mathbf{Man}$ is a diffeomorphism.

What are categories good for? 
Well, (elementary) category theory is mostly concerned with \em universal properties\em.
These define certain patterns of morphisms
that uniquely characterize (up to isomorphism) a certain mathematical structure.
An example that we will be concerned with is the notion of a `terminal object'.
Given a category $\mathbf{C}$, a \em terminal object \em is an object $I$ such that, for any object $A$ in $\mathbf{C}$, there is a unique morphism of type $f:A\to I$.
So for instance, on $\Set$ the singleton $\{ * \}$ is the terminal object, and so we obtain a characterization of the singleton set in terms of the morphisms in $\Set$.
Other standard constructions, e.g. the cartesian product, disjoint union etc. can be characterized as universal.
The power of universal properties is illustrated by their generality; e.g.~$\mathbf{Top}$ also has a terminal object, which is the one-point topological space.

Just as morphisms in a category preserve the structure of the objects, we can also define maps \em between \em categories that preserve the composition law. 
Let $\mathbf{C}$ and $\mathbf{D}$ be categories. 
A \em functor \em $F:\mathbf{C}\to\mathbf{D}$ is a mapping that: (i) assigns an object $F(A)$ in $\mathbf{D}$ to each object $A$ in $\mathbf{C}$; and (ii) assigns a morphism $F(f):F(A)\to F(B)$ to each morphism $f:A\to B$ in $\mathbf{C}$, subject to the conditions $F(g\circ f) = F(g)\circ F(f)$ and $F(1_A)=1_{F(A)}$ for all $A$ in $\mathbf{C}$.
Examples abound: we can define a powerset functor $\mathcal{P}:\Set\to\Set$ that assigns the powerset $\mathcal{P}(X)$ to each set $X$, and assigns the function $\mathcal{P}(f)::X\mapsto f[X]$ to each function $f:X\to Y$, where $f[X]\subseteq Y$ is the image of $f$.

Functors `compare' categories, and we can once again increase the level of abstraction: we can compare functors as follows.
Let $F:\mathbf{C}\to \mathbf{D}$ and $G:\mathbf{C}\to \mathbf{D}$  be a pair of functors. 
A \em natural transformation \em $\eta:F\Rightarrow G$ is a family of functions $\{\eta_A:F(A)\to G(A)\}_{A\in|\mathbf{C}|}$ indexed by the objects in $\mathbf{C}$, such that for all morphisms $f:A\to B$ in $\mathbf{C}$ we have
$ \eta_B \circ F(f) = G(f)\circ \eta_A$.

The notions of `functor' and `natural transformation' will be important in both Sections 3 and 4, when we discuss the idea of an equivalence between physical theories (characterized by means of functors), and the idea of a physical theory \textit{as} a functor respectively.

\subsection{Structuralism: simple and ontic}

As we have seen above, category theory can be thought of as `structuralist' in the following simple sense: it de-emphasizes the role played by the objects of a category, and tries to spell out as many statements as possible in terms of the morphisms between those objects. 
This much is clear from the formal properties of category theory, as evidenced by the discussion of isomorphism and universal properties above.

These formal features of category theory have developed into a vision of how to \textit{do} mathematics. This has for instance been explicitly articulated by Awodey, who says that a category-theoretic `structuralist' perspective of mathematics, is based on specifying :
\begin{quote}
 `...for a given theorem or theory only the required or relevant degree of information or structure, the essential features of a given situation, for the purpose at hand, without assuming some ultimate knowledge, specification, or ``determination'' of the objects involved.' \cite[p.~3]{Awodey2004}
 \end{quote}
Awodey presents one reasonable methodological sense of `category-theoretic structuralism': a view about how to do mathematics that is guided by the features of category theory.

Let us now contrast Awodey's sense of structuralism with a position in the philosophy of science known as Ontic Structural Realism (OSR). 
Roughly speaking, OSR is the view that the ontology of the theory under consideration is given only by structures and not by objects (where `object' is here being used in a metaphysical, and not a purely mathematical sense). Indeed, some OSR-ers would claim that:
\begin{quote}
\textbf{(Objectless)} It is coherent to have an ontology of (physical) relations without admitting an ontology of (physical) relata between which these relations hold.
\end{quote}
On the face of it, the `simple structuralism' that is evident in the practice of category theory is very different from that envisaged by OSR; and in particular, it is hardly obvious how this simple structuralism could be applied to yield (Objectless). 
On the other hand, one might venture that applying (some form of) this simple structuralism to physical theories will serve the purposes of OSR: indeed, a recent proposal by \cite{Bain2011} purports to show that the `structural methods' of category theory can be used to defend (Objectless).\footnote{Although note that Bain himself does not claim to advocate OSR.} 
Section 3 will develop and discuss Bain's suggestion, and Section 4 will clarify the category-theoretic methods on which it is based.

Before going on to evaluate this proposal, let us pause to consider which forms of OSR have an interest in such an category-theoretic argument for (Objectless). According to Frigg and Votsis' recent detailed taxonomy of structural realist positions \citep{Frigg2011}, the most radical form of OSR insists on an extensional (in the logical sense of being `uninterpreted') treatment of physical relations, i.e. physical relations are nothing but relations defined as sets of ordered tuples on appropriate formal objects. This view is faced not only with the problem of defending (Objectless) but with the further implausibility of implying that the concrete physical world is nothing but a structured set \citep[p. 261]{Frigg2011}. 

More plausible is a slightly weaker form of ontic structural realism, which Frigg calls Eliminative OSR (EOSR). Like OSR, EOSR maintains that relations are ontologically fundamental, but unlike OSR, it allows for relations that have intensions.\footnote{See \citealt[p. 262]{Frigg2011} and Ladyman and Ross for more on the distinction between OSR and EOSR. There are more sophisticated versions of EOSR---sometimes called attenuated OSR---which allow for the existence of objects but deny their individuality; we shall not consider such doctrines here. See e.g. \citealt[p. 263]{Frigg2011}.}  
Defenders of EOSR have typically responded to the charge of (Objectless)'s incoherence in various ways. For example, some claim that our ontology is `structure all the way down' without a fundamental level \citep{Ladyman2007}, or that the EOSR position should be interpreted as reconceptualizing objects as bundles of relations \citep{Morganti2004, Lam2011}.
However, Bain attempts to provide a new and novel response, i.e. a \textit{category-theoretic} defense of (Objectless). As we will see, his defense implicitly draws on two forms of category-theoretic generalization, viz. (GenDual) and (GenCat).

\section{Bain's two strategies}

Bain does not explicitly schematize the premises of his argument; nor does he divide his argument into two strategies for defending (Objectless). Thus, part of the work of this paper is to reconstruct the details of his argument with an eye towards maximal clarity and coherence. 

\cite{Bain2011} takes as his starting point the assumption that physical objects, i.e.~relata, are represented by elements of structured sets\footnote{
	In \cite[p.~12]{Bain2011}, Bain notes that if either category theory or set theory were shown to be more fundamental than the other, then this would bear on his argument for (Objectless).
	However, this issue is controversial, and our investigation will concern the extent to which category theory can be used in aid of (Objectless) \emph{without} having to broach such issues within the foundations and philosophy of mathematics.
	Hence our investigation follows Bain's lead in terms of its scope.
}, 
and proceeds to develop two argumentative strategies. 
Roughly, pp. 3-10 concern what we shall call the first strategy: it discusses (i) how set-theoretic descriptions can \textit{in general} be translated into category-theoretic descriptions, and after noting the insufficiencies of this approach, it describes (ii) a different and rather more specific sense of translation (as exemplified by the case of Einstein algebras), which is claimed to show that relata are not `essential' to the articulation of physical structure.  

On the other hand, pp. 11-12 presents a second strategy that takes it cue from examples of physical models whose very construction, it is claimed, is based on categories that do not contain relata. We will discuss each strategy in turn.

\subsection{A first strategy for defending (Objectless)}\label{sec:firstarg}

In what follows, it will be convenient to use the term `object' in two senses. First, as an object of a \textit{category}, i.e. in a purely mathematical sense.
We shall call this a \textit{$C$-object} (`C' for category-theoretic). Second, in the sense commonly used in structural realist debates, and which was already introduced above, viz. an object is a physical entity which is a relatum in physical relations. We shall call this an \textit{$O$-object} (`O' for `ontological').

We will also need to clarify our use of the term `element'. We use `element' to mean an element of a set, or as it is also often called, a `point' of a set (indeed it will be natural for us to switch to the language of points when discussing manifolds, i.e. spacetimes, in Section 4.1).
This familiar use of element should be distinguished from the category-theoretic concepts of `global element' and `generalized element', which we introduce below. 

Bain's first strategy for defending (Objectless) draws on the following idea: the usual set-theoretic representations of $O$-objects and relations can be \textit{translated} into category-theoretic terms, whence these objects can be eliminated. 
In fact, the argument can be seen as consisting of two different parts.

In the first part, Bain attempts to give a highly general argument, in the sense that it turns only on the notion of universal properties and the translatability of statements about certain mathematical representations (i.e. elements of sets) of $O$-objects into statements about morphisms between $C$-objects. 
As Bain himself notes, the general argument fails, and he thus introduces a more specific argument, which is what he wishes to endorse. The \textit{specific argument} turns on the idea of obtaining a translation scheme from a `categorical equivalence' (we explain the precise notion below in Section 4.1) between a geometric category and an algebraic category, which in turn allows one to generalize the original $C$-objects. 
The argument is `specific' because such equivalences only hold between rather special sorts of categories.

\subsubsection{The general argument}\label{sec:general}

The details of Bain's general argument can be reconstructed as follows:
\begin{quote}
$G1$: Physical objects and the structures they bear are typically identified with the elements of a set $X$ and relations on $X$ respectively.
\end{quote}

\begin{quote}
$G2$: The set-theoretic entities of $G1$ are to be represented in category-theoretic language by considering the category whose objects are the relevant structured sets, and whose morphisms are functions that preserve `structure'.\footnote{Exactly what Bain means by `structure' here will be clarified below. In particular, we will see that it is not the sense that mathematicians usually have in mind when they say that a map is `structure-preserving'.} 
\end{quote}

\begin{quote}
$G3$: Set-theoretic statements about an object of a category (of the type in $G2$) can often be expressed without making reference to the elements of that object. For instance:
\ben
\item In any category with a terminal object\footnote{See Section 2.1 for the definition of a terminal object.}
 any element of an object $X$ can be expressed as a morphism from the terminal object to $X$. (So for instance, since the singleton $\{ * \}$ is the terminal object in the category $\mathbf{Set}$, an element of a set $X$ can be described by a morphism $\{*\} \rightarrow X$.)

\item In a category with some universal property, this property can be described purely in terms of morphisms, i.e. without making any reference to elements of an object.
\een
\end{quote}

To sum up, $G1$ links $O$-objects with a standard mathematical representation, viz. elements of a set. And $G2$ and $G3$ are meant to establish the possibility that, in certain cases, category theory allows us to translate statements about elements of sets into statements about the structure of morphisms between $C$-objects. 

Thus, Bain takes $G$1--$G3$ to suggest that \cite[p.~4]{Bain2011}:
\begin{quote} 
$C$: Category theory allows for the possibility of coherently describing physical structures without making any reference to physical objects.
\end{quote}
Indeed, Bain thinks the argument suggests that the mathematical representatives of $O$-objects, i.e. the elements of sets, are \textit{surplus}, and that category theory succeeds in removing this surplus structure \citep[p. 5]{Bain2011}.\footnote{Note that even if there is surplus structure here, it is not of the same kind as, e.g. gauge-equivalent descriptions of fields in Yang-Mills theory. The latter has to do with various equivalent ways in which one can describe the dynamical objects of a theory, viz. field. By contrast, Bain's strategy involves various equivalent descriptions of the entire theory, as we will see below.} 

Bain himself thinks that the inference from $G1$--$G3$ to $C$ fails, but he does give it serious consideration, and it is easy to see why: its premises based on the most natural and general translation scheme in category theory, viz. redescribing the properties of $C$-objects in terms of morphisms, and indeed---if one is lucky---in terms of universal properties. It will be instructive for us to now discuss premises $G1$--$G3$ (which Bain endorses) in some detail as this will serve not only to highlight some technical problems with the premises, but also to draw out conceptual issues concerning Bain's interpretation of category-theoretic methods in physics, not least what he means by `structure' (and its limitations). 

First, the premise $G1$. Structural realist doctrines are typically formalized by modeling $O$-objects as elements of a set and structures as relations on that set.\footnote{There are of course exceptions: e.g. \cite{French2011a} is prepared to consider category theory as a way of formalizing structural realism if it can prove its worth.} However, this is seldom the result of reasoned deliberation about whether standard set theory is the best expressive resource from some class of such resources, but rather the product of a deeply entrenched set-theoretic viewpoint within philosophy. 
Were philosophers familiar with an alternative to set theory that was at least as powerful, e.g. category theory, then $O$-objects and structures might well have been modeled directly in the alternative formalism.\footnote{Of course, it is also a reasonable viewpoint to say that it is most `natural' to do the philosophy/foundations of physics in terms of set theory -- what is `natural' depends on how one conceives of such foundational investigations.} 

So we maintain that there is no reason for the defender of $O$-objects to accept $G1$. For instance, he might try to construct a category such that $O$-objects are modeled by $C$-objects and structures are modeled by morphisms.\footnote{
For example, there are examples of categories whose $C$-objects might coincide with the mathematical representatives of $O$-objects. For instance, in a path homotopy category, the $C$-objects are just points of the relevant space \citep{Brown2007}, and one might in turn take the points of a space to be $O$-objects, as Bain does in his example of general relativity and Einstein algebras.} 
Or he might take as his starting point a non-concrete category, whose objects have no underlying set and thus cannot be expressed in the terms of $G1$.

The premise $G2$, on the other hand, is ambiguous---it is unclear exactly how Bain wants us to understand `structure' and thus `structure-preserving maps'. First, note that when mathematicians talk about `structure-preserving maps' they usually have in mind morphisms that do not preserve all the features of a $C$-object, but rather the characteristic (albeit partial) features of that $C$-object. For instance, with respect to a group, a structure-preserving map is a \textit{homomorphism} and not an isomorphism. Bain's example of the category $\mathbf{Set}$ is of this type, because its morphisms are arbitrary functions (and not bijective functions).

However, Bain wants to introduce a different notion of `structure' that contrasts with this standard usage, for he says: 
\begin{quote}
\textbf{(Structure)}
 `...the intuitions of the ontic structural realist may be preserved by defining ``structure'' in this context to be ``object in a category''.' \citep[p.~3]{Bain2011}
 \end{quote}
 If we take this claim seriously, then a structure-preserving map will turn out to be an \textit{isomorphism} in the relevant category---for only isomorphisms preserve the complete `structural essence' of a structured set.\footnote{We thank Jonathan Bain for confirming (in private communication) that this is a fully spelled-out version of what he takes structure to be in \citep[p.~3]{Bain2011}.} 
 For instance, Bain's example of the category whose objects are smooth manifolds and whose morphisms are diffeomorphisms is of this type. If this is really what Bain has in mind, then one inevitably ends up with a very limited and dull class of categories. But even if one relaxes this notion of `structure' to mean `the structure that is preserved by the morphisms of the category, whatever they happen to be', one still runs into trouble with $G3$, as we will soon see.\footnote{Note further that, when conceiving of structured sets as objects of a (partially-specified) category, one is free to choose the morphisms (and thus fully specify the category) according to one's purpose. For instance, the categories $\Set$ and $\Rel$ have the same objects, but $\Rel$ has a much larger class of morphisms, which contain the morphisms of $\Set$ (cf. Section 3.2.2).}

We now turn to the premise $G3$. First, note that $G3$(i) is false, as we now explain. It will be convenient to introduce a piece of standard terminology: a morphism from a terminal object to some object $X$ is called a \textit{global element} of $X$. And the question of whether an element of $X$ can be expressed as a global element in the relevant category turns on the structure of the category in question. For instance, in the category $\mathbf{Man}$ with smooth manifolds as objects and smooth maps as morphisms, this question receives a positive answer: global elements are indeed in bijective correspondence with elements of a manifold.\footnote{This is because the terminal object is the $0$-dimensional manifold $\{0\}$, and so an element of a manifold $M$ is a morphism $\{0\} \rightarrow M$.}
But in many other categories, e.g. the category $\mathbf{Grp}$ defined above, the answer is negative. As an example, consider that $\mathbf{Grp}$ has the trivial group $1$ as its terminal object and so a morphism from $1$ to a group $G$ only picks out its identity and not its other elements. In order to obtain the other elements, one has to introduce the notion of a \textit{generalized element} of $X$, viz. a morphism from some `standard object' $U$ into $X$. For instance, in \textbf{Grp}, one takes $\mathbb{Z}$ as the standard object $U$, and the generalized elements $\mathbb{Z} \rightarrow G$ allow us to recover the ordinary elements of  a group $G$.  

Second, while G3(ii) is certainly true, i.e. universal properties can be expressed purely in terms of morphisms, it is a further---and significant---question for the scope and applicability of this premise whether all (or even most) physical properties can be articulated as universal properties.

Hence we have seen that the categorically-informed opponent of (Objectless) need not accept these premises---there is a lot of room for debate about how exactly one should use category theory to conceptualize the notion of physical structure. But supposing that one does: is there a valid inference from $G1$--$G3$ to $C$? Bain himself notes that the plausibility of this inference trades on an ambiguity in what one means by `reference' in $C$. If one merely means that such constructions eliminate \textit{explicit} but not implicit reference to objects, then the argument is indeed valid. On the other hand, a defense of OSR requires the elimination of \textit{implicit} reference to objects, and this is what the general argument fails to offer---it merely provides a translation scheme from statements involving elements (of sets) to statements involving morphisms between $C$-objects.
So, the defender of objects can maintain that one is still implicitly quantifying over elements. It is to overcome this objection that Bain introduces his specific argument.

\subsubsection{The specific argument}\label{sec:specific}


$G3$ above yields a special translation scheme that allows one to avoid making explicit reference to elements.
The key insight driving the specific argument is that, if one looks at a narrower range of cases, a rather different sort of translation scheme is possible: indeed one that not only avoids making explicit reference to elements, but also allows one to generalize the $C$-objects in such a way that these new $C$-objects can no longer be considered to have elements (or as many elements) in the sense of the original objects. 
According to Bain, this shows that the `...correlates [of elements of structured sets] are not \textit{essential} to the articulation of the relevant structure' \cite[p. 11]{Bain2011} (our emphasis). 
Implicit reference to elements is thereby claimed to be eliminated. 

Bain argues by appealing to a particular instance of how this translation is supposed to work, viz. the example of Einstein algebras. By contrast, we will first clarify the structure of the argument by providing its schema, and then proceed to show how Bain's example fits into this.


The starting point for Bain's specific argument is a category-theoretic version of the `semantic view of theories' on which a scientific theory is identified with its \textit{category} of models---indeed this will be the setup assumed in the following argument.\footnote{Note two points. First, we are not using `model' here in the strict sense of model theory, but rather to mean a mathematical structure that represents a physical world that is possible according to the theory. Second, this proposal is not to be confused with Lawvere's category-theoretic formulation of algebraic theories \citep{Lawvere1963}.
In the latter, models are functors between categories, whereas in the former, the models are just objects of some category (not necessarily a functor category) such as \textbf{Top}. Indeed, Lawvere's proposal is much more closely related to---though not the same as---the TQFT example that we discuss in Section 3.2. We thank an anonymous referee for pushing us to clarify this point.} 
\begin{figure}
\centering
\[
\input{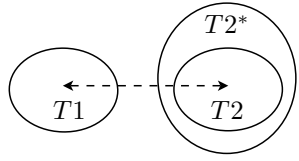}
\]
\caption{Schematic depiction of theory translation as described in $S1$--$S4$.}\label{fig:equiv}
\end{figure}
Here is our abstract reconstruction of Bain's specific argument. Let there be two theories $T_1$ and $T_2$, each represented by a category of models respectively. $T_1$ is the original physical theory that makes reference to $O$-objects.
\begin{quote}
$S1$: $T_1$ can be translated into $T_2$. In particular, each $T_1$-model can be translated into a $T_2$ model and vice versa.\end{quote}
\begin{quote}
$S2$: $T_2$ is contained in a strictly larger theory (i.e. a larger category of models) $T_2^*$. In particular, $T_2^*$ is constructed by generalizing $T_2$-models to yield models of $T_2^*$, typically by dropping an algebraic condition from the $T_2$-models. We will use $T_2'$ to denote the complement of $T_2$ in $T_2^*$.
\end{quote}
\begin{quote}
$S3$: $T_2'$ cannot be translated back into $T_1$ and so its models do not contain $T_1$-objects.
\end{quote}
\begin{quote}
$S4$: $T_2'$ is relevant for modeling some physical scenarios.
\end{quote}

When taken together, $S1$-$S4$ are supposed to show that:

\begin{quote}
$C_S$: The $T_1$-object correlates in $T_2$ do not play an essential role in articulating the physical structure (smooth structure, in Bain's specific case) of $T_{2}^*$. 
\end{quote}

Figure~\ref{fig:equiv} depicts the schema of these premises.
Let us defer for the moment the question of exactly how the idea of `translation' is supposed to work here (it is not discussed by Bain but we shall take it up in Section \ref{sec:gendual}). The key idea behind $S1$--$S4$ is that one can generalize $T_2$ to obtain a new---more general---theory $T_2^*$, some of whose models do not contain $T_1$-objects (i.e. $O$-objects in $T_1$).

In Bain's example, $T_1$ is the category of geometric models of general relativity (GR), and $T_2$ is the category of Einstein algebra (EA) models of GR (Einstein algebras were first introduced as models of GR in \citealt{Geroch1972}; we shall provide the details in Section \ref{sec:gendual}).\footnote{As we shall explain later, $T_1$ and $T_2$ are only equivalent here if we restrict the latter to the class of what we shall call \textit{Geroch} representations of EAs.} Bain is working with the idea that in geometric models of GR, the relata, or $O$-objects, of GR are space-time points of the manifold.\footnote{Note that as before (cf. G1 above), the friend of $O$-objects can challenge the idea that \textit{space-time points} should be taken to be the representatives of $O$-objects in geometric models of GR.}

We now discuss the premises of the argument and show that $S3$ rests on a technical misunderstanding; however, we will rehabilitate $S3$ before proceeding to argue that the argument fails. 
First, $S1$: Bain notes that these space-time points are in 1-1 correspondence with `maximal ideals' (an algebraic feature) in the corresponding EA model. We are thus provided with a translation scheme: points of space in a geometric description of GR are translated into maximal ideals in an algebraic description of GR. So the idea is that EA models capture the physical content of GR without making explicit reference to points.

Now the version of $S2$ that Bain uses is one in which $T_2$, the category of EAs, gets generalized to $T_2^*$, the category of sheaves of EAs over a manifold, which has a generalized notion of `smooth structure'. The former is a proper subcategory of the latter, because a sheaf of EAs over a point is just equivalent to an EA.\footnote{As we shall see later, the example of sheaves of algebraic structures is but one of a whole class of examples that could fit into this argument schema.} 

Bain then tries to obtain $S3$ by saying that a sheaf of EAs which is inequivalent to an EA does not necessarily have global elements (i.e. sections of a sheaf) in the sense previously defined, and so does not have points. Unfortunately, he confuses the notion of a local section of a sheaf of EAs (which assigns an element of an EA to an open subset of a manifold) with the notion of a maximal ideal of an EA (i.e. the algebraic correlate of a spacetime point). 
And since the two are entirely different, a lack of global sections does not imply a lack of spacetime points (i.e. $O$-objects).
Therefore $S3$ needs to be repaired.


Nonetheless, we can easily rehabilitate $S3$ is the following manner. The key idea is that while $T_1$ (a geometric model of GR) and $T_2$ (the equivalent EA model) both make reference to $T_1$-objects (explicitly and implicitly, respectively), some sheaves of EAs do not refer to $T_1$-objects because they have no formulation in terms of geometric models of GR. In other words, the \textit{generalized} smooth structure of $T_{2}'$ cannot be described in terms of the structured sets used to define \textit{ordinary} smooth structure in the case of $T_1$ and $T_2$.

Finally, as regards $S4$, various authors have taken the utility of $T'_2$ to be e.g. the inclusion of singularities in space-time \citep{Heller1992}, and as a step towards formulating quantum gravity \citep{Geroch1972}. We shall return to this point in Section 4.1.\footnote{\cite{Bain2011} does not discuss $S4$, but he does take up this matter in \citep{Bain2003}.}


We now turn to considering the inference to $C_S$. It is not entirely clear what Bain means by `[the relata] do not play an essential role'  \citep[p.~1629]{Bain2011}---nor does he expand on this phrase---but the most straightforward reading is that $T_1$-objects are eliminated \textit{simpliciter} from $T_{2}^*$.

One might compare this situation to the way that the collection of all groups (analogous to $T_2$) is contained in the collection of all monoids (analogous to $T_2^*$): it might be claimed that inverses are eliminated from the collection of all monoids. 
One could of course speak in this way, but what this would mean is that \textit{some} monoids (in particular, groups) have inverses, and some do not---a `monoid' is just a general term that covers both cases. Similarly, we can see that $C_S$ does not follow from $S1$--$S3$, since $T_2^*$ contains some models that (implicitly) quantify over $T_1$-objects, viz. the models of $T_2$, and some that do not, viz. the models of $T_2'$.\footnote{Actually, the more precise statement is that the algebraic models of $T_2'$ will not in general contain `enough' points to correspond to the geometric models of $T_1$.}  

We have seen that the specific argument will not work if one is concerned with eliminating reference to $T_1$-objects from the new and more general theory $T_{2}^*$. However, what if one is concerned not with eliminating reference, but rather with downgrading the role that $T_1$-objects play in $T_{2}^*$, e.g. by claiming that the models of $T_2'$ have a conceptual or metaphysical priority? \footnote{Indeed, Bain has (in private communication) clarified that this is how he would develop his argument in response to the above analysis.} And what would such a `downgrading' even amount to? Section 4.1 will return to these questions after a discussion of what `translation' really means in the context of $S1$.

\subsection{A second strategy for defending (Objectless)}\label{sec:secondarg}

The second strategy, as laid out in Section 4 of \cite[pp.~10--11]{Bain2011}, draws on examples of physical theories whose formulation is supposed to be essentially category-theoretic, viz. the toy models provided by $\nCob$ and $\Hilb$, and the use of these two categories to construct a Topological Quantum Field Theory (TQFT).

As before, the aim of this argument is to show that, by using category theory, physical theories can dispense with reference to physical objects, i.e.~relata (which we called $O$-objects at the start of Section \ref{sec:firstarg}).
But whereas the previous strategy is supposed to show how one can discard $O$-objects from the definition of a physical theory that is putatively formulated in terms of $O$-objects, this argument is more direct---it tells us that some category-theoretic formulations of physics (viz. $\nCob$, $\Hilb$ and TQFTs) are so \textit{general} that  they make no reference to $O$-objects to begin with, and this in turn yields (Objectless).

Here, as before, one does not need to agree with Bain's definition of an $O$-object. But even granting this definition, we shall see that the strategy is problematic. Our discussion will begin by recalling the key ideas underlying $\nCob$, $\Hilb$ and TQFTs in Section \ref{sec:pureCT}. In Section \ref{sec:OobjectsCTphysics} we distinguish the two different senses in which Bain takes $\nCob$, $\Hilb$, and TQFTs to be more general than physics based on $\Set$, after which we proceed to his claim about (Objectless).

\subsubsection{Purely category-theoretic physics?}\label{sec:pureCT}

Bain's examples in support of the second strategy are:
(i) the category $\Hilb$ of complex Hilbert spaces and linear maps; and (ii) the category $\nCob$, which has $(n-1)$-dimensional oriented closed manifolds as objects, and $n$-dimensional oriented manifolds as morphisms. 
These examples purportedly represent `purely' category-theoretic physics.
This means that formal statements about the physical theory, e.g.~quantum mechanics using $\Hilb$, are derived using the category-theoretic rules of morphisms in $\Hilb$.

Now, \textit{prima facie}, both of these examples look like good candidates for doing purely category-theoretic physics.
First, each category is potentially useful for studying the properties of quantum theory and general relativity respectively (we will elaborate further on their utility in Section \ref{sec:gencat}). Second, 
each possesses categorical properties which are promising for describing physical properties. 
More ambitiously, they suggest that one could use categorical tools to develop an approach for integrating parts of quantum theory and general relativity.

Let us pause to explain this second point, which rests on the fact that, qua categories, $\Hilb$ and $\nCob$ share some important properties.
For example, both of these categories are \em monoidal\em, meaning that both categories carry a generalisation of the tensor product $V\otimes W$ of vector spaces $V$ and $W$.\footnote{
See \cite[ch.~7]{MacLane2000} for an exposition of monoidal categories.
}
In $\nCob$ the monoidal structure is  given by the disjoint union of manifolds; whereas in $\Hilb$, the monoidal structure is given by the usual linear-algebraic tensor product of Hilbert spaces.

A second formal property shared by both categories is that they each possess a contravariant involutive endofunctor $(\cdot)^\dagger$ called the \em dagger functor\em.\footnote{
Recall that a \em contravariant \em functor is a functor $F:\mathbf{C}\to\mathbf{D}$ that reverses the direction of arrows, i.e. a morphism $f:A\to B$ is mapped to a morphism $F(f):F(B)\to F(A)$.
Also recall that an \em endofunctor \em on a category $\mathbf{C}$ is a functor $F:\mathbf{C}\to \mathbf{C}$, i.e.~the domain and codomain of $F$ are equal.} 
(This is also sometimes called a `$*$-operation', as in e.g.~\citealt{Mueger2008}.)
This means that, given a cobordism $f:A\to B$ in $\nCob$ or a linear map $L:A\to B$ in $\Hilb$, there exists a cobordism $f^\dagger:B\to A$ and a linear adjoint $L^\dagger:B\to A$ respectively, satisfying the involution laws $f^\dagger\circ f=1_A$ and $f\circ f^\dagger=1_B$, and identically for $L$.

The formal analogy between $\Hilb$ and $\nCob$ has led to the definition of a type of quantum field theory, known as a \em topological quantum field theory (TQFT)\em, first introduced by  \cite{Atiyah1988} and \cite{Witten1988}.  
A TQFT is a (symmetric monoidal) functor:
\[
T:\nCob\longrightarrow \Hilb,
\]
and the conditions placed on this functor, e.g.~that it preserve monoidal structure, reflect that its domain and target categories share formal categorical properties. 
To further flesh out the physical interpretation of TQFTs, we note that the justification for the term `quantum field theory' arises from the fact that a TQFT assigns a state space (i.e.~a Hilbert space) to each closed manifold in $\nCob$, and it assigns a linear map representing evolution to each cobordism.
This can be thought of as assigning an amplitude to each cobordism, and hence we obtain something like a quantum field theory.


\subsubsection{$O$-objects in category-theoretic physics?}\label{sec:OobjectsCTphysics}

Recall that the significance of these examples for Bain is their apparent status as purely category-theoretic formulations of physics which, in virtue of their generality, do not make any reference to $O$-objects (represented in the standard way, i.e. as elements of sets). We now turn to a criticism of this claim. 

Bain's key idea seems to be that this `generality' consists of the fact that $\nCob$ and $\Hilb$ (and thus TQFTs) have very different properties (\textit{qua} categories) from $\Set$. 
In fact, he claims that three such differences count in favor of (Objectless):
\begin{itemize}
\item[(i)] $\nCob$ and $\Hilb$ are non-concrete categories, but $\Set$ (and other categories based on it) are concrete.\footnote{Bain does not himself use the language of `non-concrete' category but this is the most reasonable---and indeed the most precise---interpretation of what he means in \cite[p. 10]{Bain2011}, since he is discussing categorical properties. At any rate, the point we make below applies equally well to his terminology of `non-structured set' without any further categorical gloss.} 
\item[(ii)] $\nCob$ and $\Hilb$ are monoidal categories, but $\Set$ is not.
\item[(iii)] $\nCob$ and $\Hilb$ have a dagger functor, but $\Set$ does not.
\end{itemize}
We address these points and their implications for (Objectless) in turn.

First, (i). Bain wants to argue that since $\nCob$ and $\Hilb$  `cannot be considered categories of structured sets' \cite[p. 10]{Bain2011}, nor can these categories be interpreted as having $O$-objects.
If one is talking about categorical properties (as in Section 4 of \cite{Bain2011}), this claim is best couched in the standard terminology that we introduced in Section \ref{sec:concrete}, viz. as the claim that these are not \textit{concrete categories}.

But this inference is faulty for two reasons. 
First, his point about non-concreteness is not altogether accurate, i.e. point (i) is false as stated.
On the one hand, it is true that $\nCob$ is not a concrete category: in particular, while the objects of $\nCob$ are structured sets, its morphisms are not functions, but manifolds, i.e.~sets equipped with the structure of a manifold.
But on the other hand, $\Hilb$ is certainly a concrete category, since the objects are Hilbert spaces, which are sets with extra conditions; and the morphisms are just functions with linearity conditions. In other words, the morphisms are structure-preserving functions. 
Thus, Bain's examples of category-theoretic physics are based in part on concrete categories.

Second and more importantly, it is doubtful that the standard mathematical notion of concreteness will aid Bain in defending (Objectless).
Bain wants to hold that the non-concreteness of a category is a sufficient condition for its not referring to $O$-objects. 
But $\nCob$ is an example of a non-concrete category that apparently contain $O$-objects---indeed the same $O$-objects (viz. space-time points) that Bain takes to be present in geometric models of GR, (cf. Section \ref{sec:general}).
We thus see that, by Bain's own lights, non-concreteness cannot be a sufficient condition of evading $O$-objects.\footnote{One could of course try to introduce the machinery of the first strategy (discussed in Section \ref{sec:specific}) to evade $O$-objects in this context, but we have already discussed the problems with this move. 
In any case, notice that the first strategy is not concerned with concreteness: translation and generalization by duality will typically yield yet another concrete category (as it does in the case of Bain's focal example, i.e.~the category of Einstein algebras, which is concrete since it is defined as a structured set).}

So the example of $\nCob$ still has $C$-objects that are based on sets, albeit morphisms which are more general than functions.
However, one can go further than this: the notion of a category is in fact defined in a schematic way (cf. Section 2.1), which leaves open the question of whether $C$-objects are sets or whether functions are morphisms. 
One might thus rhetorically ask whether this could this be the full version of `categorical generality' that Bain needs in order to defend (Objectless). 
In fact, this is implausible, because of the way in which such a schematic generality ends up being deployed in physics. As we shall see in Section 4.2, the point of schematic generality is not so that we can do physics without sets, but rather so that we can determine which categories of sets do (or do not) share important physical properties.

On to (ii): unfortunately, this claim is straightforwardly false: the category $\Set$ is certainly monoidal, with the monoidal product being given by the cartesian product (e.g.~see \citealt[p.~161]{MacLane2000}).

Finally, (iii). While it is true that $\Set$ does not have a dagger functor, and $\nCob$ and $\Hilb$ do, it is easy to construct an example of a category with a dagger functor, but which Bain would presumably agree has $O$-objects.
Consider the category $\mathbf{C}$ with one object, namely a manifold $M$ representing a relativistic spacetime; the morphisms of $\mathbf{C}$ are taken to be the automorphisms of $M$.
As with $\nCob$, this category has  natural candidates for $O$-objects (as Bain assumes), viz.~the points of the manifold.
But the category $\mathbf{C}$ also has a dagger functor: given an automorphism $f:M\to M$, the morphism $f^\dagger:M\to M$ is given by the inverse automorphism $f^{-1}$. 
In contrast, the category $\Set$ does not have a dagger functor: this follows from the observation that for any set $A$ that is not the singleton set $\{*\}$, there is a unique morphism $f:A\rightarrow \{*\}$, but the number of morphisms $g:\{*\}\rightarrow A$ is just the cardinality $| A|>1$.
Hence there does not exist a bijection between the set of morphisms $\{f:A\rightarrow \{*\}\}$ and the set of morphisms $\{g:\{*\}\rightarrow A\}$, which implies that there does not exist a dagger functor on $\Set$.
Thus, by Bain's own criterion, it is reasonable to consider $\mathbf{C}$ to be structurally dissimilar to $\Set$, despite the fact that it has $O$-objects.

More generally, i.e. putting aside the issue of (Objectless), it is quite unclear how one should interpret the physical significance of the fact that $\nCob$/$\Hilb$, but not $\Set$ has a dagger functor. For instance, it turns out that by an easy extension of $\Set$, one can construct a category that does have a dagger functor.
This easy extension is the category $\Rel$, whose objects are sets and whose morphisms are relations between objects (i.e. subsets of the Cartesian product of a pair of objects). 
Note first that $\Set$ is a subcategory of $\Rel$ because $\Set$ and $\Rel$ have same objects, and every morphism in $\Set$ is a morphism in $\Rel$.
This can be seen by noting that every function $f:A\to B$ can be written as a relation $f\subseteq A\times B$, consisting of the pairs $(a,b)$ defined by $f(a)=b$.
Second, note that---unlike $\Set$---$\Rel$ \textit{does} have a non-trivial involution endofunctor, i.e. a dagger functor, since given a relation $R:A\rightarrow B$, the relation $R^\dagger:B\rightarrow A$ is just defined by $bR^\dagger a$ if and only if $aRb$.

By extending $\Set$ to $\Rel$, we have obtained a category that is akin to $\nCob / \Hilb$ in having a dagger functor, but which also appears to be at least as good a candidate as $\Set$ for codifying the standard notion of physical structure, since its morphisms are $n$-ary relations and they can be used to encode structure. 
Evidently, if Bain's suggestion about the physical difference between these categorical properties is to amount to anything, then one must seek the right physical interpretation of such categories---a task that we shall take up in Section \ref{sec:gencat}.

We have seen that (i)--(iii) will not help Bain defend (Objectless). But what is the notion of generality that lies behind these three points? Although Bain does not state it explicitly, we can extract it from the details of his argument and summarize it as follows: 
\begin{quote}
Category theory offers us the resources to predicate properties (`categorical properties') of a category of $C$-objects without further specifying the properties of the $C$-objects, or indeed what types of objects these are (e.g. whether they are groups, manifolds, lattices, and so on). And such predications/properties are general in the sense that they are \textit{instanced} by categories of specific $C$-objects, which provide instances of such categorical properties.
\end{quote}
We call this second form of generalization (GenCat), and will explore its interpretation and use in physics (in particular the physical examples invoked by Bain) in Section \ref{sec:gencat}.

\section{Two forms of categorical generalization}\label{sec:catgen}

\subsection{Generalization by duality}\label{sec:gendual}

Earlier in Section 3.1.2, we saw that Bain invoked the notion of `translation' from geometric models of GR to EA models of GR without much explanation of its details.
As we will now see in Section 4.1.1, this idea of translation is based on the notion of an `equivalence' of categories (in contrast with the example of translation in Section \ref{sec:general}, which involved working within a category and re-describing elements of $C$-objects in terms of morphisms). We then explain the framework for this form of generalization---which we shall call \textit{generalization by duality} (GenDual)---which is based on such equivalences.  

In Section 4.1.2, we discuss the duality between geometry and algebra and then revisit Bain's EA example, which is based on such a duality. Among other things, we shall consider whether (GenDual) provides a metaphysical downgrading of $T_1$-objects---a possibility that was first raised at the end of Section 3.1.2.

\subsubsection{The framework for (GenDual)}

The relevant notion of translation is given by the notion of an \textit{equivalence} between two categories---in particular, a special kind of equivalence called a \textit{duality}.\footnote{Note that the term `duality' is used in a much looser way in the physics literature, i.e. to encompass a much more complex and profound range of examples, e.g. AdS/CFT and $S$-duality \citep{Schwarz1996}. 
Although `duality' still refers to theoretical equivalence of some kind in these examples, the nature of these equivalences is often rich and subtle, and defies formalization. On the other hand, while the category-theoretic sense of duality that we will consider is somewhat more limited in scope, it offers a precise understanding of what `theoretical equivalence' amounts to, and may in time suggest a way of understanding some of the more complex dualities that occur in physics.}
Equivalence allows us to say that two categories (and thus two theories, if we are considering categories of models) are `structurally the same'---that all the relevant content (objects, structures, operations, etc.) of each category can be described in terms of the other category. 
We will describe this notion of equivalence, explain how it underpins (GenDual), and then provide a way of spelling out the relationship between $T_1$, $T_2$, and $T_2^*$ in the specific argument, as discussed at the end of Section \ref{sec:specific}.

What does it mean to say that two mathematical objects are `structurally the same'? Earlier, we discussed the idea of isomorphism, which is a way of saying that two mathematical objects \textit{in the same category} are structurally the same. But what if the mathematical objects that we wish to discuss are themselves categories? (This is the relevant scenario when we are discussing an equivalence between two theories, whose models will generally have different types of mathematical structures respectively, and thus do not live in the same category.) A natural way in which to do so is by viewing categories themselves as the $C$-objects of a category: the category of categories, whose objects are categories and whose morphisms are functors (as defined in Section 2.1).

Hence, one might naively think that the correct way in which to articulate the structural sameness of two categories/theories is by means of the notion of an isomorphism in the category of categories.\footnote{We repeat the definition for the reader's convenience: two categories $T_1$ and $T_2$ are \textit{isomorphic} iff there exist functors $F: T_1 \rightarrow T_2$ and $G: T_2 \rightarrow T_1$ such that $G\circ F = 1_{T_1}$ and $F\circ G = 1_{T_2}$. 
}
However, this definition is too strict, as we can see from a simple example. Suppose that $T_1$ is a category with one object with only an identity morphism and no other morphisms, and that $T_2$ is a category with two isomorphic objects and two identity morphisms, and no other morphisms. Evidently, these two categories are not isomorphic. However, this conflicts with our intuition that $T_1$ and $T_2$ should really be counted as structurally `the same', because structurally speaking (i.e. up to isomorphism within a category) they contain just one object and one identity morphism. 

Fortunately, there is a standard generalization of the notion of `isomorphism of categories' that respects this intuition:
\begin{quote}
\textbf{(Equivalence)} Two categories $T_1$ and $T_2$ are \textit{equivalent} iff there exist functors $F: T_1 \rightarrow T_2$ and $G: T_2 \rightarrow T_1$ such that $G\circ F \cong 1_{T_1}$ and $F\circ G \cong 1_{T_2}$. 
 \end{quote}
where $\cong$ denotes a natural isomorphism (i.e. a natural transformation---in the sense of Section 2.1--- that is an isomorphism in the category of functors). The pair of functors $F$ and $G$ are called `quasi-inverses'. (Notice that this gives the correct result for the above `simple example', since the two objects in $T_2$ are isomorphic, but not equal.) 
A \textit{duality} is an equivalence in which $F$ and $G$ reverse the direction of the morphisms in their respective domain categories.\footnote{The important difference between (Equivalence) and isomorphism is this: the latter uses the notion of \textit{equality} to characterize $G$ and $F$ as inverses, whereas the former `relaxes' this to the notion of (natural) \textit{isomorphism}. This `relaxation' is an elementary case of what mathematicians call \textit{categorification}: a process through which one lifts sets to categories, functions to functors, and equality to isomorphism. We do not have room to discuss categorification in detail here, but suffice it to say that one can continue playing the same game for higher-order morphisms (i.e. morphisms between morphisms, and so on).}

Two equivalent categories may look rather similar, as in the simple example above. But equivalent categories can also contain objects that look very different from each other. This is often the case when, broadly speaking, one of the categories is algebraic and the other is geometric.\footnote{It is also true of many cases of theoretical equivalence in physics, where the models of the different theories often have a very different mathematical structure.}

We now have sufficient resources to make sense of mathematicians' and physicists' informal talk of `translation' in this context. Roughly speaking, if two equivalent theories are sufficiently different in the structure of their objects, then it is typical to apply the metaphor of a `language' to each category and say that one is `translating' notions from one category to the other by means of the quasi-inverse functors $F$ and $G$. In other words, what translation amounts to is that \textit{up to isomorphism}, one can construct any object in $T_1$ from an object in $T_2$ (and vice versa), and that the relational structure (i.e. the structure of the morphisms between objects) is the same in each category.\footnote{The sense of translation we have just been discussing is implemented via the quasi-inverse functors, and so applies to the objects and morphisms of each category in the pair. However, theorists also sometimes speak of translating between the bits of structure of two objects, e.g. in Bain's Einstein algebra example -- to be developed in Section 4.1.2 -- a space-time point of a manifold gets translated into a maximal ideal of an Einstein algebra. This second sense of translation depends on the construction of a particular functor between two categories, as we will see in the below examples.}

What is the significance of translation? First, it offers us two different perspectives on what is in some sense the same theory. For instance, it may be easier to prove a statement about an algebraic object than it is to prove the equivalent statement about its geometric correlate. But it also builds on this to allow for a more profound possibility, viz. a way of \textit{generalizing} a physical theory, which we call generalization by duality \textbf{(GenDual)}.

In order to discuss this profound form of generalization, we will first have to introduce a mundane form of generalization, familiar from elementary mathematics. Consider the following way of constructing a genus (category) $T^*$ of which the category $T$ is a species: we define the objects of $T^*$ by dropping some condition from the objects of $T$. So for instance, if $T$ is a category whose objects are commutative algebras, then $T^*$ could be the category of algebras, which of course contains $T$ as a subcategory.\footnote{Notice the strength of the species-genus metaphor here: the species $T$ can be defined as the conjunction of the genus $T^*$ and some differentia (i.e. the commutativity of the algebra, in this case).}

The profound generalization (GenDual) is achieved when one applies the following procedure:

\begin{enumerate}

\item \textit{Equivalence}: Let $T_1$ and $T_2$ be equivalent (perhaps dual) theories.

\item \em Mundane Generalization\em: Let $T_2$ be generalized to a larger theory $T_{2}^*$ by dropping some differentia from the definition of the objects of $T_2$.

\item \textit{Analogy}: By combining Equivalence and Mundane Generalization, one finds that there is a sense in which $T_{2}'$-objects (i.e. the complement of $T_2$ in $T_{2}^*$) can be thought of as generalized versions of $T_1$-objects.This sense will not be literal, since $T_{2}'$-objects cannot be translated (via equivalence) into $T_1$-objects. However, Equivalence allows us to obtain a characterization of $T_1$-objects in terms of $T_2$, and if the essential features of this characterization are preserved in the objects of $T_{2}'$, then one can call the latter `$T_1$-objects' by analogy. 
\end{enumerate}

An example of this procedure (which we discuss at greater length below) is where $T_1$ is a category of geometries and $T_2$ is a category of algebras. Mathematicians typically deform $T_2$ to a category $T_2*$ which relaxes some condition on the objects of $T_2$ (e.g. the existence of sufficient points) but which retains algebraic analogs of geometric features (e.g. symmetries, invariants, open sets, measures) and operations (integration and differentiation) of $T_1$. The objects of $T_2'$ are then called `generalized geometries' by Analogy.\footnote{What are the essential (algebraic correlates of geometric) features whose retention in $T_2'$ suffices for $T_2'$-objects to be analogically called `geometries'? We do not pretend to give a metaphysical answer to this question. Rather, our answer comes from the practice of mathematics: mathematicians take certain features (e.g. open sets satisfying certain axioms) as being definitive of certain objects (e.g. a topology) and if one can hang on to the algebraic correlates of these notions while eliminating the correlates of other features (e.g. points) then one has preserved the `essential features'.}  

Returning now to the main topic of this section, viz. (GenDual), we can now use the idea of \emph{core-dependent homonymy} \citep{Shields1999} to explain the mathematical practice of describing $T_2'$-objects as $T_1$-objects even when they are evidently not the same thing: generalization by duality establishes $T_1$ (which is equivalent to $T_2$) as the primary case of things called $T_{2}^*$, and so $T_2'$ is related to $T_2$ as a secondary case is to a primary case. In other words, $T_2$ is not just a species of $T_2^*$ in this scenario, but rather an `essential' or `core' species of $T_2^*$. 

It follows that the form of generality given by (GenDual) is such that one cannot define $T_2$ as a genus (e.g. $T_2^*$) in conjunction with some differentia---this would leave out the fact that $T_2$ is the essential species of the genus, and that all the other species of $T_2^*$ are to be interpreted with respect to $T_2$.  

An example will serve to make this reasoning more concrete: it will be the instance of algebraic-geometric duality that is implicitly appealed to in Bain's specific argument. We now turn to discussing this duality.

\subsubsection{Algebraic-geometric duality and Einstein algebras}\label{sec:algebraic}

The duality most commonly invoked in order to accomplish (GenDual) is called `algebraic-geometric duality', where `algebra' is understood as algebraic structures induced by classes of functions on a space, and `geometry' is understood broadly to include topological spaces and various kinds of geometric structures that can be added to a topological space (e.g. vector bundles with connection, spin structures, geometric operators acting on the space, etc.).

One of the simplest examples of algebraic-geometric duality is the duality between $T_1$, the (geometric) category of locally compact Hausdorff topological spaces, and $T_2$, the category of commutative $C^*$-algebras.\footnote{$T_1$ has continuous proper maps as morphisms, and $T_2$ has proper $*$-homomorphisms as morphisms.}

The quasi-inverse (and contravariant) functors $F: T_1 \rightarrow T_2$ and $G: T_2 \rightarrow T_1$ are easy to describe. Roughly speaking, $F$ takes a topological space to the commutative $C^*$-algebra of complex-valued continuous functions on that space. On the other hand, $G$ (also called the `functor of points') takes each algebra to the set of characters (i.e. homomorphisms from the algebra to $\mathbb{C}$) of the algebra. These characters are then thought of as points of a topological space, and when they are suitably topologized, one recovers the entire structure of the relevant topological space.\footnote{Here we describe only the action of the functors on objects; both $F$ and $G$ act on morphisms by pullbacks, see e.g. Section 1 of \cite{Khalkali2009}.}  It is then a straightforward matter to check that $T_1$ is equivalent to $T_2$, i.e. that $F$ and $G$ satisfy the natural isomorphisms:
\begin{equation}
GF \cong 1_{T_1}  \qquad FG \cong 1_{T_2}.
\end{equation}

What are the implications of this duality for translating concepts between $T_1$ and $T_2$? First, two algebras in $T_2$ are isomorphic (as algebras) if and only if their spaces of characters (given by action of $G$) are homeomorphic (which is what the notion of isomorphism amounts to in $T_1$), and vice versa.\footnote{Similarly, the group of automorphisms of an algebra in $T_1$ is isomorphic to the group of homeomorphisms of the character space of that algebra.} Furthermore, there is a correspondence between the topological properties of a space in $T_1$ and the properties of an algebra in $T_2$. For instance, compact spaces correspond to unital $C^*$-algebras, an open subset of a space corresponds to an ideal of an algebra, a closed subset of a space corresponds to a quotient algebra, a measure on a space corresponds to a positive functional on an algebra, and so on. Thus topological and algebraic `correlates' can be translated back and forth between $T_1$ and $T_2$.

The next step of (GenDual), viz. \textit{Mundane Generalization} can be implemented by extending $T_2$ to the category $T_2^*$ of 
\em all \em $C^*$-algebras. In particular, the complement $T_2'$ is the category of non-commutative $C^*$-algebras. 

The last step of (GenDual), viz. \textit{Analogy}, turns on a happy but contingent fact about this combination of Equivalence and Mundane Generalization, viz. that the algebraic objects in the complement $T_2'$ still have most of the algebraic properties that, when possessed by objects of $T_2$, corresponded to topological properties of spaces via equivalence. Indeed, there is only one significant property that separates $T_2'$ from $T_2$: dropping the commutativity condition from $C^*$-algebras makes it the case that these algebras no longer have a rich set of characters (i.e. the correlates of points of a topological space); in particular they will no longer correspond to objects in the category of topological spaces. Nonetheless, because the other algebraic correlates in $T_2$ of topological concepts in $T_1$ still make sense in $T_2'$, mathematicians exploit this analogy between $T_2$ and $T_2'$ and deem the objects of $T_2'$ to be non-commutative `topological spaces'. This practice is reminiscent of the phenomenon of core-dependent homonymy, where there can be e.g. two non-univocal terms -- indeed homonyms -- one of which provides the core sense of the term, whereas the other term has a meaning that is derived from the core sense. So, in the present example of (GenDual), the core sense of (locally compact Hausdorff) topological spaces is provided by the category $T_1$ and its equivalent category of algebras $T_2$, and the non-commutative `topological spaces' of $T_2'$ derive their meaning from this equivalence.\footnote{More sophisticated equivalences of algebraic-geometric duality abound, and they extend various geometrical concepts in a similar way, e.g. integration, vector fields, metrics, topological invariants, and symmetries of such spaces, to name just a few examples. Many of these dualities also have physical applications, e.g. Alain Connes' use of non-commutative geometry to attempt  to describe various quantum field theories and the Higgs mechanism \citep{Connes2008}.}

Now that we have a better understanding of the foundation of Bain's first strategy, viz. (GenDual), we are ready to revisit his Einstein algebra (EA) example from the perspective of algebraic-geometric duality. This will also provide us with the occasion to analyze the utility of (GenDual) within a physical context.

In his paper, Bain refers primarily to \citeapos{Heller1992} formulation of EAs. But in order to understand the physical interpretation of (GenDual), it will be helpful to start by considering Geroch's motivations when he first introduced EAs in \citep{Geroch1972}. Geroch's original motivation was to find a description of GR that might be useful for extending it to describe quantum gravity. In particular, Geroch wanted to accommodate the following plausible intuition about a quantum theory of gravity \citep[p.~1]{Geroch1972}: 
\begin{quote} `...it is perhaps reasonable to expect that, in a quantum theory of gravitation, the mathematical formalism will, at some point, suggest a ``smearing out of events'''.
\end{quote}
Thus, he sought a description of models of GR that would eliminate explicit reference to space-time events, i.e. points of the space-time manifold. In order to do so, he exploited a familiar fact from differential geometry: the vector fields on a manifold can be characterized in purely algebraic terms, i.e. as the set of derivations acting on the algebra of smooth real-valued functions.

Geroch developed this familiar fact into a purely algebraic analogue of the theory of tensors, and proceeded to introduce algebraic versions of the metric, the covariant derivative, and various tensor fields---thus arriving at an algebraic model of Einstein's equations and its solutions, which he called Einstein algebras (EAs).\footnote{Roughly speaking, Geroch defined an EA as an algebra $F$, with a subring $R$ that is isomorphic to the real numbers, such that we can define an $F$-module (corresponding to the set of vector fields) which satisfies all the necessary algebraic conditions for it to count as a model of GR.} 

Geroch showed that every geometric model of GR gives rise to an EA. In other words, he defined a functor $F$ from $T_1$, the category of geometric models of GR, to a subcategory of the category of all EA models. However, he did not try to formulate an \textit{equivalence} between $T_1$ and the category of EA models.\footnote{We thank Robert Geroch for clarifying this with us in private communication.} Indeed, \cite{Geroch1972} does not use the language of category theory at all.

A more sophisticated development of EAs was then accomplished by \cite{Heller1992}, who showed that $T_1$ is equivalent, in fact \textit{dual}, to $T_2$, the category of (what Heller called) `Geroch representations' of EAs. Similarly to our previous example of the duality between topological spaces and commutative $C^*$-algebras, Heller introduced a functor of points $G$ in order to build a space from an Einstein algebra and to play the role of a quasi-inverse with respect to the functor $F$. Indeed, his functor provides even more information because it associates to such a space a class of functions by means of a `functional representation' of the EA. 

It will now be helpful to elucidate Heller's work by means of our explanation of (GenDual) above. First, he showed that there was an Equivalence (indeed a duality) between $T_1$, the category of geometric models of GR, and $T_2$, the category of Geroch representations of EAs, whose class of functions is isomorphic to  $C^\infty (M)$, i.e. the set of smooth functions on a smooth manifold $M$. Second, by Mundane Generalization, he extended $T_2$ to the category $T_2^*$ of general EA representations. The important thing about this extension is that $T_2'$ contains objects that need not be smooth manifolds, and whose class of functions is \textit{not} isomorphic to $C^\infty (M)$, but are nonetheless still equipped with an algebraic notion of `global smooth structure' which only coincides with geometric notion of smoothness (associated with $C^\infty (M)$) when one restricts to the sub-category of Geroch representations, i.e. $T_2$. Thus, the objects of $T_2'$ cannot be translated into the geometric objects of $T_1$, whose notion of smoothness is inherently bound up with the idea of local functions. Nonetheless, since the objects of $T_2'$ still have an algebraic notion of smoothness and all the algebraic operations on `algebraically smooth' objects that are necessary for formulating GR, we can---along with Heller---use Analogy to consider these objects to be `generalized models of GR'.

Note that Heller's motivation for extending $T_2$ to $T_2^*$ was not in order to describe quantum gravity (at least in the first instance) but rather in order to describe singularities as part of the structure of EA models of GR. By contrast, in traditional geometric GR, singularities are defined by the inextendibility of causal curves in the space-time, and as such are not part of the intrinsic structure of space-time. Ordinary smooth $n$-manifolds have trouble accommodating singularities precisely because of the point-like nature of how their `smoothness' is defined. Heller's extended category $T_2^*$ overcomes this problem because the objects of $T_2'$ have no such local smoothness condition. \cite{Heller1992} then went on to generalize this construction further, by introducing sheaves of Einstein algebras over a differential space, which allows for the inclusion of an even wider class of singularities than that described by $T_2'$---we shall however leave this aside, as the present sketch is sufficient both to illustrate a physical application of (GenDual), as well as to further discuss Bain's argument in light of it.

Now that we have gained more perspective on the application of (GenDual) to Bain's EA example, we can revisit the issue that we left open at the end of Section 3.1, viz. whether---instead of claiming that (GenDual) for EAs eliminates $T_1$-objects---one can claim that it merely downgrades the metaphysical status of $T_1$-objects, because it gives metaphysical priority to the models of $T_2'$.

Note that, at least from the purely mathematical point of view, the above analysis suggests that it is the models of $T_1 / T_2$ that have conceptual priority (and the models of $T_2'$ that are derivatively called `generalized models of $T_1$). Nonetheless, one might try to deploy the following analogy in a physical context.
Perhaps one should think of the new (generalized) models in $T_{2}^{'}$ as standing in relation to $T_2$ in the way that quantum mechanics---when formalized in terms of non-commutative $C^*$-algebras---stands in relation to classical mechanics, which can be formalized in terms of commutative $C^*$-algebras.\footnote{See e.g. \cite{Rieffel1994}'s seminar paper on deformation quantization for a review of how we can conceptualize classical and quantum mechanics in this way.}
In this analogy, $T_1$ is the classical phase space and $T_2$ is its dual algebra of functions on that phase space, i.e. a commutative $C^*$-algebra. One can then extend $T_2$ to $T_2^*$ via deformation quantization \citep{Rieffel1994}, and this new category will include the non-commutative $C^*$-algebras which are required for quantum mechanics. The complement of $T_2$, again denoted $T'_2$, can be thought of as playing the role of a generalized `non-commutative' phase space.
Thus, one might try to say that since quantum mechanics is metaphysically prior to classical mechanics, the $O$-objects of classical phase space (whatever they are) are `ontologically downgraded' in $T_2^*$, even if they are not eliminated.


Unfortunately, this analogy is unsatisfactory for at least two reasons. First, while the non-commutative deformation of classical phase space constitutes a complete and adequate successor theory, $T'_2$ for EAs does not. Recalling Geroch and Heller's motivations for constructing an EA description of GR (which we have just discussed) will highlight this point. From Geroch's perspective, viz. trying to construct a theory of quantum gravity, $T'_2$ is merely a \textit{suggestive template} (in the sense that it does include point-events) with which one might begin to construct a theory of quantum gravity. From Heller's perspective, $T'_2$ is just an extension of the models of $T_2$ that is \textit{perhaps} useful in gaining physical insight into singularities (it is at the very least unclear whether it really provides additional insight, since there are standard ways of dealing with singularities even with respect to geometric models of GR). 
Thus, it seems clear that (GenDual) leaves open the question of how the models of $T_2'$ are to be interpreted with respect to those of $T_1 / T_2$.

Second, in the case of EAs, it is implausible to view $T_2$ as being recovered from $T_2^*$ in any \textit{physical} sense. This is because the models of $T_2$ are `recovered' from $T_2^*$ by simply \textit{adding} some algebraic condition to $T_2^*$ models (e.g. the defining condition of the Geroch representation of EAs). By contrast, the commutative algebras of classical mechanics are recovered from the non-commutative algebras of quantum mechanics by taking the limit of a physical parameter $\hbar \rightarrow 0$. The point here is that, even if one could attribute physical significance to the defining algebraic conditions of $T_2'$ and $T_2$ (e.g. spacetimes that do and do not include an intrinsic description of singularities respectively) one still needs to be able to give a robust account of why $T_2'$ should be viewed as the metaphysical core of $T_2^*$, and how $T_2$ can be recovered from $T_2^*$ by means of a physical limit.

In any case, these considerations show that even if one manages to obtain a theory $T_2^*$ in which the models of $T_1$ (and thus $T_1$-objects) are metaphysically downgraded, this downgrading will not be accomplished by (GenDual), but rather by the physical interpretation of the models of $T_2'$---which turn on details which are manifestly independent of (GenDual).\footnote{\textit{A fortiori}, it is not category theory which does the work of accomplishing this downgrading, since category theory is crucial only in establishing the \textit{equivalence} step of (GenDual).} 

Let us now summarize this Section and generalize from the example of EAs to draw a positive moral for the physical application of (GenDual). Our discussion has shown that Bain's first strategy to defend (Objectless) fails. Our discussion of his general argument showed the limitations of trying to translate statements about elements (of set-theoretic objects) into statements that involve only morphisms of a category. Our discussion of his specific argument showed that it is underwritten by (GenDual). Furthermore, we argued that while (GenDual) -- and the associated use of category theory -- does not necessarily have any bearing on ontology, it is a useful methodological tool for generating new physical theories. Indeed, one might say that it provides the following recipe for constructing or \textit{partially} constructing novel theories (or extensions of theories):
\begin{enumerate}
\item Collect some initial class of models into a category $T_1$.
\item Identify some feature $f$ that you would like the new models to exhibit, but which is inconsistent with $T_1$. 
\item Implement (GenDual), i.e. find a category $T_2$ that is equivalent to $T_1$, and extend it to $T_2^*$ by deforming $T_2$ to possess (the translated version of) $f$. By core-dependent homonymy, $T_2^*$ can be called the `generalized version' of $T_1$. 
\end{enumerate}  

Let us again emphasize that this recipe cannot on its own determine the interpretive content of the novel models, e.g. whether they should be thought of as `fringe cases' of an extended version of $T_2$, or as the core content of a new and more fundamental theory, from which $T_2$ is to be recovered under some appropriate physical conditions. And as we saw in the example of Geroch's motivation for developing EAs (i.e. in order to describe quantum gravity), the recipe will in general suggest nothing more than a template for the latter purpose. One will often have to do further, conceptually independent work in order to fully specify such novel models and explain how they effectively yield the models of $T_2$ under the appropriate conditions.

\subsection{Generalization by categorification}\label{sec:gencat}

We now turn to the task of developing the Bain's observations about structural differences between categories.
In particular, we shall discuss (i) how Bain's examples of structural differences between categories leads us to another form of generalization, viz. (GenCat); and (ii) the example of TQFTs, which is complex and subtle -- in particular, it is not clear what work (if any) (GenCat) is doing in TQFTs.

\subsubsection{The framework for (GenCat)}

In Section \ref{sec:OobjectsCTphysics}, we argued that formulating physics in categories which are structurally different from $\Set$ would not help Bain's defense of (Objectless). However, it is apparent that Bain's second strategy has uncovered a phenomenon of great interest, viz. that one can generalize the properties of physical theories from the level of individual sets to the level of abstract categorical properties, and that such properties often have interesting physical interpretations and implications. The abstract categorical properties, e.g. that of a category's being symmetric monoidal, find \textit{instances} in particular physical categories, e.g. $\Hilb$ or $\nCob$.
We shall call this \emph{generalization by categorification}, or (GenCat) for short.  

Before providing a precise framework for describing this sequence of steps, let us explain how it is motivated by the structural differences that Bain discusses.
For example, consider 
the categories $\Set$ and $\Hilb$.
Both categories are \emph{monoidal}, meaning that, given any two objects $A$ and $B$ in the category, there exists a third object $A\otimes B$.\footnote{
The physical interpretation of the monoidal product requires a general interpretation of the category---this is indeed clear in the case of $\Hilb$, since it describes some of the mathematics of quantum mechanics.
See \citep{Coecke2009d} for an interpretation of the category $\Set$.
}
If one interprets the objects of the category as abstract state spaces (as is indeed suggested by $\Hilb$), then one can interpret the monoidal product as capturing the concept of `composition of state spaces'.

Although $\Set$ and $\Hilb$ are both instances of monoidal categories, they each have a very different kind of monoidal structure, which indicates a structural difference between these categories.
On the one hand, in $\Set$ the monoidal structure is given by the cartesian product, and so an element of a composite `system' $A\times B$ is always a pair $(a,b)$.
If we consider $\Set$ to represent physical systems, then this corresponds to the fact that the `parts determine the whole'---a feature that we would associate with classical systems.
On the other hand, in $\Hilb$ the monoidal structure is given by the tensor product of Hilbert spaces, for which there are elements of $\mathcal{H}_1\otimes \mathcal{H}_2$ that are \em not \em of the form $\psi_1\otimes\psi_2$, i.e.~there exist \em entangled states\em.

We can thus \emph{identify} a physical property, viz.~entanglement, that is possessed by monoidal categories such as $\Hilb$.
However, the above way of identifying this property relies on a distinction that is based on how the objects in each category are \emph{internally} defined -- in other words, the identification is based on a property of the elements of the (set-based) objects of the category.

This set-theoretic distinction is hardly transparent in terms of the overall structure of $\Set$ and $\Hilb$.
However, it turns out one can achieve such transparency---and indeed generality---by formulating the distinction in terms of category theory.
For a monoidal category $\mathbf{C}$, the absence of entanglement in the above sense corresponds to $\mathbf{C}$ satisfying the axioms of a \emph{cartesian category}, of which $\Set$ is an instance.\footnote{A cartesian category is a monoidal category supplemented with an extra condition, viz. that for any object $A$,  there are projection morphisms $\pi_1$ and $\pi_2$ that guarantee that every element $p:I\to A$ satisfies $p=\langle\pi_1\circ p,\pi_2\circ p\rangle$.
The category $\Set$ is an example, since for any element $(a,b)\in X$ there projection morphisms $\pi_1\circ(a,b)=a$ and $\pi_2\circ(a,b)=b$.}
On the other hand, a category that is monoidal but not cartesian can be said to possess the relevant notion of entanglement.
We thus see that given an appropriate background interpretation, the notion of a cartesian category provides a category-theoretic distinction between theories with a notion of entanglement, and those without.\footnote{
	The label of `categorification' is also used
	by Baez in \cite[p. 12]{Baez2004}.
	Our usage is similar but subtly different. 
	In particular, the programme of categorification pursued by Baez typically includes further steps such as replacing identity statements with isomorphisms.
}

Moreover, we have achieved the following generality in our description: since this notion of entanglement has been described categorically, it can be applied to any monoidal category (whose objects are interpreted as state spaces and whose morphisms are interpreted as dynamical processes) and not just $\Hilb$ and $\Set$.
Take for instance the category $\Rel$ (introduced in Section \ref{sec:OobjectsCTphysics}), which has the same objects as $\Set$, but which is not a cartesian category.
One might thus say that $\Rel$ can be considered to `have entanglement'---in order for this locution to be meaningful, of course, we must be able to interpret $\Rel$ physically, e.g. as in \citep{Edwards2009,Coecke2009}, where the morphisms are interpreted as non-deterministic classical processes between state spaces.

Let us take stock: the physical applications that we have described---and on which Bain draws---turn out to have a significance that is rather different from Bain's morals.
In particular, the generality implicit in formulating structural differences (say between some category $\mathcal{C}$ and $\Set$) at a categorical level does not suggest (Objectless), but rather that one can use such categorical properties to qualitatively distinguish between theories, each of which is defined as a category of structured sets.

The key steps behind this generalization by categorification (GenCat) can be articulated as follows:
\begin{enumerate}
	\item \emph{Identification:} Isolate a physical phenomenon in a specific category $\mathbf{C}$ of set-based objects;
	\item \emph{Categorical Definition:} Define the phenomenon in purely categorical terms, i.e. without referring to any specific category.
	\item \emph{Application:} Apply this definition to other specific categories, which will instantiate the physical phenomenon.
\end{enumerate}

The utility of (GenCat), is three-fold.
Firstly, as we saw above with the example of $\Rel$ and $\Hilb$, (GenCat) can be used to \emph{classify} theories (thought of as categories) according to the type of physical resources that they have.
Secondly, (GenCat) can be used to generate new models with the physical properties that are desired.
For example, given the category-theoretic definiton of entanglement, any monoidal category that is not cartesian can potentially serve as a representation of a physical theory with entanglement. 
Thirdly, this allows for an incisive high-level analysis of the logical relationship between different physical phenomenona: a striking example is given in \cite{Coecke2009,Coecke2011a}, where it is shown---using categorical definitions---that there exist categories which have entanglement and complementary observables but which do not provide non-locality.
Thus non-locality is shown to be independent of entanglement.

Before concluding this section, let us briefly return to Bain's argument concerning (Objectless). 
Recall that Bain's thinks of physical structure as something that is borne by a $C$-object in a category (cf. (Structure) in Section 3.1.1).
On the other hand, our discussion above show that the categorical properties of $\Hilb$ can not only be used to formalize the physics of quantum systems, but also that doing so encodes physical structure in the algebra of morphisms of the relevant category (which does not correspond to either the structure of the objects, or the specific morphisms which preserve the object's structure). 
This provides a further reason for thinking either that Bain's notion of physical structure is too limited, and that his second strategy is not relevant to evaluating the plausibility of (Objectless).

\subsubsection{A closer look at TQFTs}\label{sec:tqfts}

We have used $\Hilb$ to illustrate (GenCat), but Bain also discusses another interesting example, viz. topological quantum field theories (TQFTs), which build on his two examples of symmetric monoidal categories ($\Hilb$ and $\nCob$).
This raises the question: does the use of TQFTs in physics exemplify (GenCat)?
As we shall see in the following, the answer is rather subtle.

We will focus on understanding the significance of a result that we shall refer to as `Kock's theorem' below.
This result is of interest for two reasons: first, it is a central result in the the study of TQFTs; second, it appears to be an instance of (GenCat), at least \emph{prima facie}.

Recall that a TQFT is a functor:
\[
T:\nCob\longrightarrow \Hilb.
\]
Since $T$ is a functor, it consists of an assignment of objects and morphisms to the category $\nCob$.
First, $T$ assigns a quantum state space, viz.~a Hilbert space, to an object  in $\nCob$.
And second, $T$ assigns a linear map to a morphism in $\nCob$, i.e.~a cobordism.
Since a cobordism is a a kind of `spacetime', the linear map that $T$ assigns can be thought of as an evolution operation. 
As we discussed in Section \ref{sec:secondarg}, if $n=2$, then the objects in $\nCob$ are disjoint unions of circles, and the morphisms are 2-dimensional oriented topological spaces, where each of these has a boundary given by a disjoint union of circles.

The spacetime structure in $\mathbf{2Cob}$ is evidently rather simple.
In particular, a TQFT is a \em topological \em theory. 
This means that quantities that are calculated for a particular type of evolution are the same for any topological deformation of that evolution.
For example, the probability amplitude for transition from one state to another
depends only on the topology change that a cobordism supplies.
So a TQFT will assign the same linear operator in $\Hilb$ to the following two morphisms in $\nCob$:
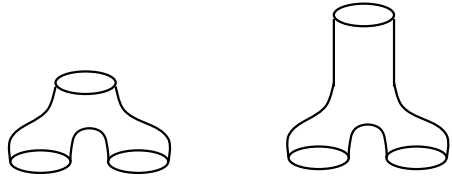
\[
\ifx\JPicScale\undefined\def\JPicScale{1}\fi
\psset{unit=\JPicScale mm}
\psset{linewidth=0.3,dotsep=1,hatchwidth=0.3,hatchsep=1.5,shadowsize=1,dimen=middle}
\psset{dotsize=0.7 2.5,dotscale=1 1,fillcolor=black}
\psset{arrowsize=1 2,arrowlength=1,arrowinset=0.25,tbarsize=0.7 5,bracketlength=0.15,rbracketlength=0.15}
\begin{pspicture}(0,0)(61.48,24)
\rput{0}(7,3){\psellipse[](0,0)(4,-1.5)}
\rput{0}(13,13.5){\psellipse[](0,0)(4,-1.5)}
\rput{0}(20,3){\psellipse[](0,0)(4,-1.5)}
\psecurve(3,3)(3,3)(3,6)(8,10)(9,13)(9,13)(4,17)
\psecurve(24,3)(24,3)(24,6)(18,10)(17,13)(17,13)(12,17)
\psecurve(11,3)(11,3)(11,4)(12,7)(15,7)(16,4)(16,3)(16,3)
\rput{0}(44,3.5){\psellipse[](0,0)(4,-1.5)}
\rput{0}(50,22.5){\psellipse[](0,0)(4,-1.5)}
\rput{0}(57,3.5){\psellipse[](0,0)(4,-1.5)}
\psecurve(40,3.5)(40,3.5)(40,6.5)(45,10.5)(46,13.5)(46,13.5)(41,17.5)
\psecurve(61,3.5)(61,3.5)(61,6.5)(55,10.5)(54,13.5)(54,13.5)(49,17.5)
\psecurve(48,3.5)(48,3.5)(48,4.5)(49,7.5)(52,7.5)(53,4.5)(53,3.5)(53,3.5)
\psline(54,22)(54,13)
\psline(46,22)(46,13)
\end{pspicture}

\]

A particularly interesting feature of TQFTs for our purposes is that $2$-dimensional TQFTs can be classified in a purely category-theoretic way.
To explain the classification theorem, it will be helpful to introduce the graphical calculus that exists for describing TQFTs. 

For example, consider the cobordisms:
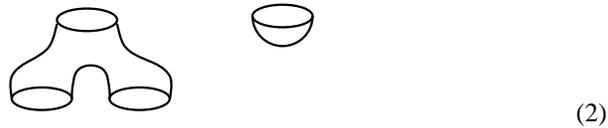
\beq\label{eq:merge}
\ifx\JPicScale\undefined\def\JPicScale{1}\fi
\psset{unit=\JPicScale mm}
\psset{linewidth=0.3,dotsep=1,hatchwidth=0.3,hatchsep=1.5,shadowsize=1,dimen=middle}
\psset{dotsize=0.7 2.5,dotscale=1 1,fillcolor=black}
\psset{arrowsize=1 2,arrowlength=1,arrowinset=0.25,tbarsize=0.7 5,bracketlength=0.15,rbracketlength=0.15}
\begin{pspicture}(0,0)(43,18)
\rput{0}(7,3){\psellipse[](0,0)(4,-1.5)}
\rput{0}(13,13.5){\psellipse[](0,0)(4,-1.5)}
\rput{0}(20,3){\psellipse[](0,0)(4,-1.5)}
\psecurve(3,3)(3,3)(3,6)(8,10)(9,13)(9,13)(4,17)
\psecurve(24,3)(24,3)(24,6)(18,10)(17,13)(17,13)(12,17)
\psecurve(11,3)(11,3)(11,4)(12,7)(15,7)(16,4)(16,3)(16,3)
\rput{0}(39,14){\psellipse[](0,0)(4,-1.5)}
\rput{0}(39,14){\psellipticarc[](0,0)(4,4){180}{360}}
\end{pspicture}

\eeq
A TQFT detects only topology change,  e.g.~the `merging' of two circles to one circle in Eq.~\ref{eq:merge}.
We can depict this as:
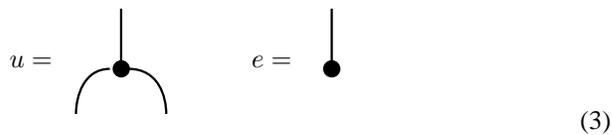
\beq\label{eq:frob}
\ifx\JPicScale\undefined\def\JPicScale{1}\fi
\psset{unit=\JPicScale mm}
\psset{linewidth=0.3,dotsep=1,hatchwidth=0.3,hatchsep=1.5,shadowsize=1,dimen=middle}
\psset{dotsize=0.7 2.5,dotscale=1 1,fillcolor=black}
\psset{arrowsize=1 2,arrowlength=1,arrowinset=0.25,tbarsize=0.7 5,bracketlength=0.15,rbracketlength=0.15}
\begin{pspicture}(0,0)(49,16)
\psbezier(21,8)(25,8)(26,4.29)(26,2)
\psbezier(18.5,8)(15,8)(14,4.29)(14,2)
\rput(8,9){$u=$}
\psline(20,16)(20,8.62)
\rput{0}(20,8){\psellipse[fillstyle=solid](0,0)(1,1)}
\rput(40,9){$e=$}
\psline(48,16)(48,8.25)
\rput{0}(48,8){\psellipse[fillstyle=solid](0,0)(1,1)}
\end{pspicture}

\eeq
Similarly, the associativity of the `merge' operation represented by the cobordism in Eq.~\ref{eq:merge} is represented as:
\beq\label{eq:frob2}
\ifx\JPicScale\undefined\def\JPicScale{1}\fi
\psset{unit=\JPicScale mm}
\psset{linewidth=0.3,dotsep=1,hatchwidth=0.3,hatchsep=1.5,shadowsize=1,dimen=middle}
\psset{dotsize=0.7 2.5,dotscale=1 1,fillcolor=black}
\psset{arrowsize=1 2,arrowlength=1,arrowinset=0.25,tbarsize=0.7 5,bracketlength=0.15,rbracketlength=0.15}
\begin{pspicture}(0,0)(49,22)
\psbezier(8,6.48)(12,6.48)(13,3.09)(13,1)
\psbezier(5.5,6.48)(2,6.48)(1,3.09)(1,1)
\psline(7,11.04)(7,7.04)
\rput{0}(7,6.48){\psellipse[fillstyle=solid](0,0)(1,0.91)}
\psbezier(14,15.61)(18,15.61)(19,12.22)(19,10.13)
\psbezier(11.5,15.61)(8,15.61)(7,12.22)(7,10.13)
\psline(13,22)(13,16.17)
\rput{0}(13,15.61){\psellipse[fillstyle=solid](0,0)(1,0.91)}
\psline(19,10.13)(19,1)
\rput(25,10.13){$=$}
\psbezier(38,15.61)(42,15.61)(43,12.22)(43,10.13)
\psbezier(35.5,15.61)(32,15.61)(31,12.22)(31,10.13)
\psline(37,22)(37,16.17)
\rput{0}(37,15.61){\psellipse[fillstyle=solid](0,0)(1,0.91)}
\psbezier(44,6.48)(48,6.48)(49,3.09)(49,1)
\psbezier(41.5,6.48)(38,6.48)(37,3.09)(37,1)
\psline(43,11.04)(43,7.04)
\rput{0}(43,6.48){\psellipse[fillstyle=solid](0,0)(1,0.91)}
\psline(31,10.13)(31,1)
\end{pspicture}

\eeq
Now, Eq.~\ref{eq:frob} and Eq.~\ref{eq:frob2}  can be considered to be \em graphical laws\em, in the sense that the topological deformations can now be seen as rules for rewriting graphs (see \citealt{Selinger2011a} for a more formal discussion of graphical languages in monoidal categories).
These graphical laws correspond to algebraic laws, and thus: the diagrams in Eq.~\ref{eq:frob} correspond to morphisms $u:A\otimes A\rightarrow A$ and $e:A\rightarrow I$, and the graphical rules in Eq.~\ref{eq:frob2} define a \em monoid object \em on the object $A$.

Now, as we have stated it, Eq.~\ref{eq:frob2} is an equation in $\nCob$, since the morphisms $u$ and $e$ are cobordisms.
But we can define morphisms $u$ and $e$, and impose Eq.~\ref{eq:frob2} in \emph{any} monoidal category $\mathbf{C}$.
In particular, we can define a monoid object in the category $\Set$: in this case a monoid object is just a monoid---i.e.~the usual notion of a set with a binary operation and a unit element.
This process of abstraction can be thought of as \em categorifying \em the cobordism laws (or indeed the rules for a monoid), since it amounts to defining categorical rules for the topological laws of cobordisms.
Note also that we can continue to use the graphical rules in $\mathbf{C}$, since these correspond to the algebraic rules.
The definition of a monoid object here looks like an instance of (GenCat)---to what extent is this true?
We shall return to this question shortly, but note that (GenCat) requires the identification of a physical phenomenon before the categorification step.


Given our graphical notation, we can define a similar set of graphical rules to Eq.~\ref{eq:frob2} (see, e.g.~\citealt[p.~87]{Coecke2009d}), but with the diagrams reflected in the $x$-axis. 
For example, given morphisms $\delta:A\rightarrow A\otimes A$ and $\epsilon:A\rightarrow I$, we can consider the condition:
\beq\label{eq:frob3}
\ifx\JPicScale\undefined\def\JPicScale{1}\fi
\psset{unit=\JPicScale mm}
\psset{linewidth=0.3,dotsep=1,hatchwidth=0.3,hatchsep=1.5,shadowsize=1,dimen=middle}
\psset{dotsize=0.7 2.5,dotscale=1 1,fillcolor=black}
\psset{arrowsize=1 2,arrowlength=1,arrowinset=0.25,tbarsize=0.7 5,bracketlength=0.15,rbracketlength=0.15}
\begin{pspicture}(0,0)(49,23.84)
\psbezier(8,17.93)(12,17.93)(13,21.59)(13,23.84)
\psbezier(5.5,17.93)(2,17.93)(1,21.59)(1,23.84)
\psline(7,13.02)(7,17.33)
\rput{0}(7,17.93){\psellipse[fillstyle=solid](0,0)(1,-0.98)}
\psbezier(14,8.09)(18,8.09)(19,11.75)(19,14)
\psbezier(11.5,8.09)(8,8.09)(7,11.75)(7,14)
\psline(13,1.2)(13,7.49)
\rput{0}(13,8.09){\psellipse[fillstyle=solid](0,0)(1,-0.98)}
\psline(19,14)(19,23.84)
\rput(25,14){$=$}
\psbezier(38,8.09)(42,8.09)(43,11.75)(43,14)
\psbezier(35.5,8.09)(32,8.09)(31,11.75)(31,14)
\psline(37,1.2)(37,7.49)
\rput{0}(37,8.09){\psellipse[fillstyle=solid](0,0)(1,-0.98)}
\psbezier(44,17.93)(48,17.93)(49,21.59)(49,23.84)
\psbezier(41.5,17.93)(38,17.93)(37,21.59)(37,23.84)
\psline(43,13.02)(43,17.33)
\rput{0}(43,17.93){\psellipse[fillstyle=solid](0,0)(1,-0.98)}
\psline(31,14)(31,23.84)
\end{pspicture}

\eeq
These rules define a \em comonoid\em, where the prefix `co' is used to indicate that the direction of the arrows---or, correspondingly, the orientation of the diagram---is reversed.
If an object $A$ carries both a monoid and a comonoid (subject to compatibility laws, known as the \em Frobenius  laws\em), then $A$ is said to carry a \em Frobenius algebra\em.


Now, to state the classification theorem we must define the category of two-dimensional TQFTs, denoted $\mathbf{2TQFT}$: its objects are TQFTs, i.e. the functors $T$ defined above, and its morphisms are monoidal natural transformations.
There is then an algebraic classification of TQFTs as follows.
Consider the category $\mathbf{cFA}$ of commutative Frobenius algebras and Frobenius algebra homomorphisms. 
As \cite{Kock2003} explains, a TQFT induces a Frobenius algebra:
a TQFT $F:\nCob\to\Hilb$ 
preserves monoidal structure, and so it preserves the Frobenius algebra laws such as  Eq.~\ref{eq:frob2} and Eq.~\ref{eq:frob3}.
Hence a TQFT defines a commutative Frobenius algebra in $\Hilb$.
Conversely, it can be shown that a Frobenius algebra defines a TQFT.
Thus we obtain the following equivalence, due in part to \cite{Abrams1996} (and with much useful exposition by \citealt{Kock2003}):
\[
\mathbf{2TQFT} \simeq \mathbf{cFA}.
\]
We refer to this equivalence as \em Kock's theorem\em\footnote{
This name is justified by the fact that, although the roots of this result lie with the other authors that we have cited, \cite{Kock2003} is the main source for the explicitly category-theoretic formulation of this result.
}.

Now, recall that Bain introduced TQFTs as an example of category-theoretic generalization that is similar to $\Hilb$.
However, having sketched the details of this generalization, viz. (GenCat), we can see that they are in fact rather different.
On the one hand, $\Hilb$ is an instance of a symmetric monoidal category, and by interpreting its morphisms as the dynamical processes of quantum mechanics, we can formulate various physical properties (e.g. entanglement) at the level of general categorical properties.
On the other hand, a TQFT is a functor between two specific symmetric monoidal categories (viz. $\nCob$ and $\Hilb$), and it is not obvious to see how one can abstractly formulate any information of physical relevance at the level of categorical properties.
Kock's theorem is important precisely because it provides a sense in which such an abstraction occurs, viz. it says that a TQFT is equivalent to a Frobenius algebra object, which can be defined in any symmetric monoidal category.

Can Kock's theorem then be described as an example of (GenCat)? 
Prima facie, the answer is not clear, because it is unclear which physical properties are being encoded in the categorical structure of `Frobenius algebra object in a category'; and it is also unclear which specific categories can provide physical instantiations of this structure.

However, a partial answer is forthcoming from recent developments in physics. 
For instance, it has been shown that the algebraic sector of some conformal field theories can be classified by certain Frobenius algebra objects \citep{Runkel2007}. 
In this case, the Frobenius algebra encodes the fundamental (algebraic) structure of the theory, and it is instantiated in the category of representations of different conformal field theories.
Our conceptual work shows one way in which the elaboration of this work can be construed as an instance of (GenCat): the `physical phenomenon' that is being identified is the abstract structure of a conformal field theory, which is then described in `purely categorical terms' as a Frobenius algebra object. By choosing the category in which this Frobenius algebra object lives, one then arrives at a determinate instance of this abstract structure, viz. a particular conformal field theory.

\subsection{The r\^ole of category theory in physics}\label{sec:CT}

Thus far, we have identified two forms of category-theoretic generalization that are employed in physics.
We will now comment on some differences between the two strategies, and emphasise the significance of these differences.


In the first type of generalization, viz. (GenDual), the set-theoretic models of a theory are simply collected into a category---it thus follows that these models (and thus the theory) are not defined categorically, i.e. their mathematical structure does not rely only (or even primarily) on category theory.
On the other hand, the second form of generalization, viz. (GenCat), does indeed aspire to define (the abstraction of) a physical phenomenon in purely categorical terms.

For this reason, one sees that morphisms play a different role in (GenDual) and (GenCat) respectively. 
In (GenDual), morphisms are typically taken to be the automorphisms of a model (e.g. the diffeomorphisms from a spacetime to itself, in Bain's example) and thus do not connect different objects in a category.
But in (GenCat), the relevant physical phenomena need to be represented in way that is germane to category-theoretic abstraction, e.g. dynamical processes are typically represented as morphisms between objects, and composition is represented by a functor. The generalization then comes about when we ask that these structures retain their meaning in virtue of the abstract properties of the relevant category, e.g. being symmetric monoidal and having a dagger functor. 

These considerations point to a central difference in the role that category theory plays for (GenDual) and (GenCat) respectively.
In (GenDual), the main role of category theory is to provide a way of saying when two theories (represented by categories) are equivalent; the generalization then comes about by extending one of these theories and using analogy to conceive of the new objects in the extension as `generalized models' of the other theory. Thus, while category theory provides the right language to express `equivalence', which is in turn necessary (but insufficient) for constructing the analogy, it is clear that it does not itself provide the desired generalization; nor is this form of generalization `native' to category theory.

By contrast, in (GenCat), it is precisely category theory which accomplishes the desired generalization, since this is brought about by reformulating the physically relevant properties of some specific category such as $\Hilb$ as the properties of an abstract category, whose objects are not further specified as being of any particular type. This physical application of category theory is the analog of taking the notion of a set-theoretical mathematical object, e.g. a monoid, and generalizing it to a purely category-theoretic concept, e.g. a monoid object, which can be formulated in any monoidal category.

The differences between the two strategies can be summarised as follows.
We might say that in the first strategy, category theory is playing an \emph{organizational} role, in the sense that it is employed to collect together the models of a theory in a single mathematical structure, i.e.~a category.
The morphisms in this category are `scaffolding' for the other mathematics that is employed, and do not play a role in elucidating the physical concepts. 
On the other hand, the second strategy exhibits the \emph{direct} use of category theory in physics.
Here, category theory is used to directly encode physical processes.
Among other things, this means that the choice of underlying category is potentially not crucial for doing calculations: what's important is that it satisfies certain category-theoretic axioms (e.g. having monoidal structure), which are then used to compute physical consequences.



\section{Conclusion}


Our path has taken us from Bain's argument for OSR to uncovering two paradigmatic forms of generalization for physical theories.
Our work suggests two specific avenues for further research.

Firstly, it raises the question of the extent to which the forms of reasoning that we have uncovered rely on category theory \emph{per se}.
That is, could generalization by duality and generalization by categorification be implemented with something other than the mathematics of category theory?
This might seem rather unlikely in the latter case, but a priori it could be that, e.g.~lattice theory, could play an equally powerful r\^ole in formalizing the types of generalization.

Furthermore, one might observe that the use of category theory in physics is a relatively recent development---and thereby argue that its prominence could be faddish.
However, there are reasons to think otherwise.
For example, although the initial work on TQFTs in \citep{Witten1988} was not itself explicitly category-theoretic, much of the current work in this field now uses category theory in a widespread and deep way, as shown by the work of \cite{Baez1995} and \cite{Runkel2007}.
This suggests a kind of evolution in the definition of a physical theory: from less category-theoretic formalisms to more category-theoretic formalisms. 
If such a process holds true in general, then it could point to the essential utility of category theory in physics.


Secondly, it is interesting to note that  \cite{Awodey2004} has advocated that category theory provides a formalism for one sense of mathematical structuralism or what he calls `schematism'.
Part of his argument concerns distinguishing between the methodological and metaphysical significance of the use of category theory in pure mathematics.
Since this has been one of our concerns in this paper---but with respect to the use of category theory in physics---the work of Awodey is ripe for comparison with ours.

\paragraph{Acknowledgements.}

\bibliographystyle{apalike}
\bibliography{refs}

\begin{thebibliography}{}

\bibitem[Abrams, 1996]{Abrams1996}
Abrams, L. (1996).
\newblock Two-dimensional topological quantum field theories and {F}robenius
  algebras.
\newblock {\em Journal of Knot Theory and its Ramifications}, 5(05):569--587.

\bibitem[Atiyah, 1988]{Atiyah1988}
Atiyah, M. (1988).
\newblock Topological quantum field theories.
\newblock {\em Publications Math{\'e}matiques de l'IH{\'E}S}, 68(1):175--186.

\bibitem[Awodey, 2004]{Awodey2004}
Awodey, S. (2004).
\newblock An answer to {H}ellman's question: Does category theory provide a
  framework for mathematical structuralism?'.
\newblock {\em Philosophia Mathematica}, 12(1):54--64.

\bibitem[Baez, 2004]{Baez2004}
Baez, J. (2004).
\newblock Quantum quandaries: a category-theoretic perspective.
\newblock arXiv:quant-ph/0404040.

\bibitem[Baez and Dolan, 1995]{Baez1995}
Baez, J.~C. and Dolan, J. (1995).
\newblock Higher-dimensional algebra and topological quantum field theory.
\newblock {\em Journal of Mathematical Physics}, 36:6073.

\bibitem[Bain, 2003]{Bain2003}
Bain, J. (2003).
\newblock Einstein algebras and the hole argument.
\newblock {\em Philosophy of Science}, 70 (5):1073--1085.

\bibitem[Bain, 2013]{Bain2011}
Bain, J. (2013).
\newblock Category-theoretic structure and radical ontic structural realism.
\newblock {\em Synthese}, 190:1621--1635.

\bibitem[Brown, 2007]{Brown2007}
Brown, R. (2007).
\newblock Topology and groupoids.
\newblock {\em Bull. London Math. Soc}, 39:867--872.

\bibitem[Coecke et~al., 2009]{Coecke2009}
Coecke, B., Edwards, B., and Spekkens, R. (2009).
\newblock The group theoretic origin of non-locality for qubits.
\newblock Technical Report RR-09-04, OUCL.

\bibitem[Coecke et~al., 2011]{Coecke2011a}
Coecke, B., Edwards, B., and Spekkens, R.~W. (2011).
\newblock Phase groups and the origin of non-locality for qubits.
\newblock {\em Electronic Notes in Theoretical Computer Science}, 270(2):15 --
  36.
\newblock arXiv:1003.5005.

\bibitem[Coecke and Paquette, 2009]{Coecke2009d}
Coecke, B. and Paquette, E.~O. (2009).
\newblock Categories for the practising physicist.
\newblock arXiv:0905.3010.

\bibitem[Connes and Marcolli, 2008]{Connes2008}
Connes, A. and Marcolli, M. (2008).
\newblock {\em Noncommutative geometry, quantum fields and motives}, volume~55.
\newblock American Mathematical Soc.

\bibitem[Edwards, 2009]{Edwards2009}
Edwards, B. (2009).
\newblock {\em Non-locality in categorical quantum mechanics}.
\newblock PhD thesis, University of Oxford.

\bibitem[Esfeld and Lam, 2011]{Lam2011}
Esfeld, M. and Lam, V. (2011).
\newblock Ontic structural realism as a metaphysics of objects.
\newblock In Bokulich, A. and Bokulich, P., editors, {\em Scientific
  structuralism}, page 143–159. Dordrecht: Springer.

\bibitem[French, 2011]{French2011a}
French, S. (2011).
\newblock Shifting to structures in physics and biology: A prophylactic for
  promiscuous realism.
\newblock {\em Studies in History and Philosophy of Science Part C: Studies in
  History and Philosophy of Biological and Biomedical Sciences},
  42(2):164--173.

\bibitem[Frigg and Votsis, 2011]{Frigg2011}
Frigg, R. and Votsis, I. (2011).
\newblock Everything you always wanted to know about structural realism but
  were afraid to ask.
\newblock {\em European journal for philosophy of science}, 1(2):227--276.

\bibitem[Geroch, 1972]{Geroch1972}
Geroch, R. (1972).
\newblock Einstein algebras.
\newblock {\em Communications in Mathematical Physics}, 26(4):271--275.

\bibitem[Heller, 1992]{Heller1992}
Heller, M. (1992).
\newblock Einstein algebras and general relativity.
\newblock {\em International journal of theoretical physics}, 31(2):277--288.

\bibitem[Khalkali, 2009]{Khalkali2009}
Khalkali, M. (2009).
\newblock {\em Basic Non-commutative Geometry}.
\newblock EMS.

\bibitem[Kock, 2003]{Kock2003}
Kock, J. (2003).
\newblock {\em Frobenius Algebras and 2-D Topological Quantum Field Theories}.
\newblock London Mathematical Society Student Texts. Cambridge University
  Press.

\bibitem[Ladyman and Ross, 2007]{Ladyman2007}
Ladyman, J. and Ross, D. (2007).
\newblock {\em Every thing must go: Metaphysics naturalized}, volume~61.
\newblock Oxford University Press Oxford.

\bibitem[{Lam} and {Wuthrich}, 2013]{Lam2013}
{Lam}, V. and {Wuthrich}, C. (2013).
\newblock {No categorial support for radical ontic structural realism}.
\newblock {\em ArXiv e-prints}.

\bibitem[Lambek and Scott, 1988]{Lambek1988}
Lambek, J. and Scott, P. (1988).
\newblock {\em Introduction to Higher-Order Categorical Logic}.
\newblock Cambridge Studies in Advanced Mathematics. Cambridge University
  Press.

\bibitem[Landry, 2013]{Landry2013}
Landry, E. (2013).
\newblock The genetic versus the axiomatic method: Responding to feferman 1977.
\newblock {\em The Review of Symbolic Logic}, 6:24--51.

\bibitem[Lawvere, 1963]{Lawvere1963}
Lawvere, F.~W. (1963).
\newblock Functorial semantics of algebraic theories.
\newblock {\em Proceedings of the National Academy of Sciences of the United
  States of America}, 50(5):869.

\bibitem[{Mac Lane}, 2000]{MacLane2000}
{Mac Lane}, S. (2000).
\newblock {\em Categories for the Working Mathematician}.
\newblock Springer-Verlag.

\bibitem[Morganti, 2004]{Morganti2004}
Morganti, M. (2004).
\newblock On the preferability of epistemic structural realism.
\newblock {\em Synthese}, 142(1):81--107.

\bibitem[M\"uger, 2008]{Mueger2008}
M\"uger, M. (2008).
\newblock Tensor categories: A selective guided tour.
\newblock arXiv:0804.3587.

\bibitem[Rieffel, 1994]{Rieffel1994}
Rieffel, M. (1994).
\newblock Quantization and $c^*$ algebras.
\newblock {\em Contemporary Mathematics}, 167:67--97.

\bibitem[Runkel et~al., 2007]{Runkel2007}
Runkel, I., Fjelstad, J., Fuchs, J., and Schweigert, C. (2007).
\newblock Topological and conformal field theory as frobenius algebras.
\newblock {\em Contemporary Mathematics}, 431:225.

\bibitem[Schwarz, 1997]{Schwarz1996}
Schwarz, J.~H. (1997).
\newblock {Lectures on superstring and M theory dualities: Given at ICTP Spring
  School and at TASI Summer School}.
\newblock {\em Nucl.Phys.Proc.Suppl.}, 55B:1--32.

\bibitem[Selinger, 2011]{Selinger2011a}
Selinger, P. (2011).
\newblock A survey of graphical languages for monoidal categories.
\newblock In Coecke, B., editor, {\em New Structures for Physics}, volume 813
  of {\em Lecture Notes in Physics}, pages 289--355. Springer Berlin /
  Heidelberg.

\bibitem[Shields, 1999]{Shields1999}
Shields, C. (1999).
\newblock {\em Order in Multiplicity: Homonymy in the Philosophy of Aristotle}.
\newblock Oxford University Press on Demand.

\bibitem[Witten, 1988]{Witten1988}
Witten, E. (1988).
\newblock Topological quantum field theory.
\newblock {\em Communications in Mathematical Physics}, 117(3):353--386.

\end{thebibliography}

\end{document}